\batchmode
\makeatletter
\def\input@path{{"/home/jacob/Documents/Work/My Papers/2023-Stochastic Processes and Quantum Theory/"}}
\makeatother
\documentclass[11pt,twoside,english]{article}
\usepackage[T1]{fontenc}
\usepackage[latin9]{inputenc}
\usepackage{babel}
\usepackage{mathtools}
\usepackage{amsmath}
\usepackage{amssymb}
\usepackage{geometry}
\usepackage{setspace}
\usepackage[pdftex,pdfusetitle,
 bookmarks=false,
 breaklinks=false,pdfborder={0 0 0},pdfborderstyle={},backref=false,colorlinks=false]
 {hyperref}

\makeatletter

\usepackage{endnotes}

\let\originalleft\left
\let\originalright\right
\renewcommand{\left}{\mathopen{}\mathclose\bgroup\originalleft}
\renewcommand{\right}{\aftergroup\egroup\originalright}

\def\smalloverbrace#1{\mathop{\vbox{\m@th\ialign{##\crcr%
      \noalign{\kern3\p@}%
      \tiny\downbracefill\crcr\noalign{\kern3\p@\nointerlineskip}%
      $\hfil\displaystyle{#1}\hfil$\crcr}}}\limits}

\def\smallunderbrace#1{\mathop{\vtop{\m@th\ialign{##\crcr
   $\hfil\displaystyle{#1}\hfil$\crcr
   \noalign{\kern3\p@\nointerlineskip}%
   \tiny\upbracefill\crcr\noalign{\kern3\p@}}}}\limits}

\DeclareMathAlphabet{\mymathbb}{U}{bbold}{m}{n}

\makeatother

\begin{document}
\title{The Stochastic-Quantum Correspondence}
\author{Jacob A. Barandes\thanks{Departments of Philosophy and Physics, Harvard University, Cambridge, MA 02138; jacob\_barandes@harvard.edu}
\thanks{This manuscript is the accepted version, as published in the journal \emph{Philosophy of Physics} (DOI:10.31389/pop.186). Some material that was removed since Version 1 has been incorporated into arxiv:2507.21192.} }
\date{June 30, 2025}
\maketitle
\begin{abstract}
This paper argues that every quantum system can be understood as a
sufficiently general kind of stochastic process unfolding in an old-fashioned
configuration space according to ordinary notions of probability.
This argument is based on an exact correspondence between the class
of \textquoteleft indivisible\textquoteright{} stochastic processes
and quantum theory. This new stochastic-quantum correspondence demotes
the wave function from a primary ontological ingredient to a secondary
mathematical tool, and yields a deflationary account of exotic quantum
phenomena, such asinterference, decoherence, entanglement, noncommutative
observables, and wave-function collapse. At a more practical level,
the stochastic-quantum correspondence leads to a novel reconstruction
of quantum theory, alongside the Hilbert-space, path-integral, and
quasiprobability representations, and also provides a framework for
using Hilbert-space methods to formulate highly generic, non-Markovian
types of stochastic dynamics, with potential applications throughout
the sciences.
\end{abstract}

\begin{center}
\global\long\def\quote#1{``#1"}%
\global\long\def\apostrophe{\textrm{'}}%
\global\long\def\slot{\phantom{x}}%
\global\long\def\eval#1{\left.#1\right\vert }%
\global\long\def\keyeq#1{\boxed{#1}}%
\global\long\def\importanteq#1{\boxed{\boxed{#1}}}%
\global\long\def\given{\vert}%
\global\long\def\mapping#1#2#3{#1:#2\to#3}%
\global\long\def\composition{\circ}%
\global\long\def\set#1{\left\{  #1\right\}  }%
\global\long\def\setindexed#1#2{\left\{  #1\right\}  _{#2}}%

\global\long\def\setbuild#1#2{\left\{  \left.\!#1\,\right|\,#2\right\}  }%
\global\long\def\complem{\mathrm{c}}%

\global\long\def\union{\cup}%
\global\long\def\intersection{\cap}%
\global\long\def\cartesianprod{\times}%
\global\long\def\disjointunion{\sqcup}%

\global\long\def\isomorphic{\cong}%

\global\long\def\setsize#1{\left|#1\right|}%
\global\long\def\defeq{\equiv}%
\global\long\def\conj{\ast}%
\global\long\def\overconj#1{\overline{#1}}%
\global\long\def\re{\mathrm{Re\,}}%
\global\long\def\im{\mathrm{Im\,}}%

\global\long\def\transp{\mathrm{T}}%
\global\long\def\tr{\mathrm{tr}}%
\global\long\def\adj{\dagger}%
\global\long\def\diag#1{\mathrm{diag}\left(#1\right)}%
\global\long\def\dotprod{\cdot}%
\global\long\def\crossprod{\times}%
\global\long\def\Probability#1{\mathrm{Prob}\left(#1\right)}%
\global\long\def\Amplitude#1{\mathrm{Amp}\left(#1\right)}%
\global\long\def\cov{\mathrm{cov}}%
\global\long\def\corr{\mathrm{corr}}%

\global\long\def\absval#1{\left\vert #1\right\vert }%
\global\long\def\expectval#1{\left\langle #1\right\rangle }%
\global\long\def\op#1{\hat{#1}}%

\global\long\def\bra#1{\left\langle #1\right|}%
\global\long\def\ket#1{\left|#1\right\rangle }%
\global\long\def\braket#1#2{\left\langle \left.\!#1\right|#2\right\rangle }%

\global\long\def\parens#1{(#1)}%
\global\long\def\bigparens#1{\big(#1\big)}%
\global\long\def\Bigparens#1{\Big(#1\Big)}%
\global\long\def\biggparens#1{\bigg(#1\bigg)}%
\global\long\def\Biggparens#1{\Bigg(#1\Bigg)}%
\global\long\def\bracks#1{[#1]}%
\global\long\def\bigbracks#1{\big[#1\big]}%
\global\long\def\Bigbracks#1{\Big[#1\Big]}%
\global\long\def\biggbracks#1{\bigg[#1\bigg]}%
\global\long\def\Biggbracks#1{\Bigg[#1\Bigg]}%
\global\long\def\curlies#1{\{#1\}}%
\global\long\def\bigcurlies#1{\big\{#1\big\}}%
\global\long\def\Bigcurlies#1{\Big\{#1\Big\}}%
\global\long\def\biggcurlies#1{\bigg\{#1\bigg\}}%
\global\long\def\Biggcurlies#1{\Bigg\{#1\Bigg\}}%
\global\long\def\verts#1{\vert#1\vert}%
\global\long\def\bigverts#1{\big\vert#1\big\vert}%
\global\long\def\Bigverts#1{\Big\vert#1\Big\vert}%
\global\long\def\biggverts#1{\bigg\vert#1\bigg\vert}%
\global\long\def\Biggverts#1{\Bigg\vert#1\Bigg\vert}%
\global\long\def\Verts#1{\Vert#1\Vert}%
\global\long\def\bigVerts#1{\big\Vert#1\big\Vert}%
\global\long\def\BigVerts#1{\Big\Vert#1\Big\Vert}%
\global\long\def\biggVerts#1{\bigg\Vert#1\bigg\Vert}%
\global\long\def\BiggVerts#1{\Bigg\Vert#1\Bigg\Vert}%
\global\long\def\ket#1{\vert#1\rangle}%
\global\long\def\bigket#1{\big\vert#1\big\rangle}%
\global\long\def\Bigket#1{\Big\vert#1\Big\rangle}%
\global\long\def\biggket#1{\bigg\vert#1\bigg\rangle}%
\global\long\def\Biggket#1{\Bigg\vert#1\Bigg\rangle}%
\global\long\def\bra#1{\langle#1\vert}%
\global\long\def\bigbra#1{\big\langle#1\big\vert}%
\global\long\def\Bigbra#1{\Big\langle#1\Big\vert}%
\global\long\def\biggbra#1{\bigg\langle#1\bigg\vert}%
\global\long\def\Biggbra#1{\Bigg\langle#1\Bigg\vert}%
\global\long\def\braket#1#2{\langle#1\vert#2\rangle}%
\global\long\def\bigbraket#1#2{\big\langle#1\big\vert#2\big\rangle}%
\global\long\def\Bigbraket#1#2{\Big\langle#1\Big\vert#2\Big\rangle}%
\global\long\def\biggbraket#1#2{\bigg\langle#1\bigg\vert#2\bigg\rangle}%
\global\long\def\Biggbraket#1#2{\Bigg\langle#1\Bigg\vert#2\Bigg\rangle}%
\global\long\def\angs#1{\langle#1\rangle}%
\global\long\def\bigangs#1{\big\langle#1\big\rangle}%
\global\long\def\Bigangs#1{\Big\langle#1\Big\rangle}%
\global\long\def\biggangs#1{\bigg\langle#1\bigg\rangle}%
\global\long\def\Biggangs#1{\Bigg\langle#1\Bigg\rangle}%

\global\long\def\vec#1{\mathbf{#1}}%
\global\long\def\vecgreek#1{\boldsymbol{#1}}%
\global\long\def\idmatrix{\mymathbb{1}}%
\global\long\def\projector{P}%
\global\long\def\permutationmatrix{\Sigma}%
\global\long\def\densitymatrix{\rho}%
\global\long\def\krausmatrix{K}%
\global\long\def\stochasticmatrix{\Gamma}%
\global\long\def\lindbladmatrix{L}%
\global\long\def\dynop{\Theta}%
\global\long\def\timeevop{U}%
\global\long\def\hadamardprod{\odot}%
\global\long\def\tensorprod{\otimes}%

\global\long\def\inprod#1#2{\left\langle #1,#2\right\rangle }%
\global\long\def\normket#1{\left\Vert #1\right\Vert }%
\global\long\def\hilbspace{\mathcal{H}}%
\global\long\def\samplespace{\Omega}%
\global\long\def\configspace{\mathcal{C}}%
\global\long\def\phasespace{\mathcal{P}}%
\global\long\def\spectrum{\sigma}%
\global\long\def\restrict#1#2{\left.#1\right\vert _{#2}}%
\global\long\def\from{\leftarrow}%
\global\long\def\statemap{\omega}%
\global\long\def\degangle#1{#1^{\circ}}%
\global\long\def\trivialvector{\tilde{v}}%
\global\long\def\eqsbrace#1{\left.#1\qquad\right\}  }%
\par\end{center}

\section{Introduction\label{sec:Introduction}}

The theory of stochastic processes describes the phenomenological
behavior of systems with definite configurations that evolve probabilistically
in time. Quantum theory is a comprehensive mathematical apparatus
for making measurement predictions when taking into account the microscopic
constituents of various kinds of physical systems, from subatomic
particles to superconductors. At an empirical level, both theories
involve probabilities, and at the level of formalism, both employ
vectors and matrices.

The primary goal of this paper is to introduce an exact correspondence
between a highly general class of stochastic processes and quantum
theory, within which measuring devices and observers are incorporated
as ordinary subsystems. This \emph{stochastic-quantum correspondence}
takes the form of a simple \textquoteleft dictionary\textquoteright{}
expressing any time-dependent stochastic matrix in terms of a suitable
combination of Hilbert-space ingredients.

From a practical standpoint, the stochastic-quantum correspondence
provides a systematic framework for constructing highly generic forms
of stochastic dynamics, much as the Lagrangian and Hamiltonian formulations
of classical mechanics provide systematic frameworks for constructing
deterministic dynamics. Potential applications range from turbulence
to finance, to name just two examples. Importantly, the stochastic-quantum
correspondence does not require assuming that the stochastic dynamics
in question can be modeled as a Markov chain. 

Taking a more foundational perspective, this paper also uses the
stochastic-quantum correspondence to show that physical models based
on old-fashioned configuration spaces and ordinary forms of probability,
combined with stochastic dynamics, can replicate all the empirical
predictions of textbook quantum theory\textemdash including interference,
decoherence, entanglement, noncommutative observables, and wave-function
collapse\textemdash without relying on the austere and metaphysically
opaque Dirac-von Neumann axioms (Dirac 1930, von Neumann 1932)\nocite{Dirac:1930pofm,vonNeumann:1932mgdq}.
In this alternative approach, a given system moves stochastically
along a physical trajectory in a prosaic, classical-looking configuration
space.  The ingredients of the Hilbert-space formulation, including
the wave function, then go the way of the luminiferous aether of 19-century
electromagnetism\textemdash they are no longer primary or ontological
features of the theory.\footnote{For a discussion of some of the outstanding problems in the philosophy
of quantum theory, see Myrvold (2022)\nocite{Myrvold:2022piiqt}.
For an extensive analysis of the role of the wave function in quantum
theory, see Ney, Albert (2013)\nocite{NeyAlbert:2013twfeotmoqm}.}

At the very least, this approach yields a new formulation of quantum
theory, one that is based on a picture of stochastic systems evolving
in configuration spaces within the framework of ordinary probability
theory. This formulation therefore joins a list of ways to formulate
quantum systems that include the traditional Hilbert-space formulation
(Dirac 1930, von Neumann 1932)\nocite{Dirac:1930pofm,vonNeumann:1932mgdq},
the path-integral formulation (Dirac 1933, Feynman 1942, Feynman 1948)\nocite{Dirac:1933tliqm,Feynman:1942tpolaiqm,Feynman:1948statnrqm},
and the quasi-probability formulation (Wigner 1932, Moyal 1949)\nocite{Wigner:1932otqcfte,Moyal:1949qmaast}.
As noted by Feynman (1948)\nocite{Feynman:1948statnrqm}, ``there
is a pleasure in recognizing old things from a new point of view,''
and ``there is always the hope that the new point of view will inspire
an idea for the modification of present theories, a modification necessary
to encompass present experiments.''

The present work is not continuous with earlier efforts to identify
a fundamental relationship that connects  stochastic processes and
quantum theory. The most well-known of these approaches are due to
Bopp~(1947, 1952, 1953)\nocite{Bopp:1947qsuk,Bopp:1952efdqbsdk,Bopp:1953sudgdqde},
Fényes (1952)\nocite{Fenyes:1952ewbuidq}, and Nelson (1967, 1985)\nocite{Nelson:1966dtobm,Nelson:1985qf}.
For a contemporary example, see Friederich (2024)\nocite{Friederich:2024itqbioqt}.
Altogether different are stochastic-collapse models (Ghirardi, Remini,
Weber 1986; Bassi, Ghirardi 2003)\nocite{GhirardiRiminiWeber:1986udmms,BassiGhirardi:2003drm},
in which a quantum system's wave function or density matrix is assumed
to experience stochastic fluctuations through time. 

Section~\ref{sec:Indivisible-Stochastic-Processes} will start with
the definition of an \emph{indivisible stochastic process}, along
with introducing the key distinction between \emph{divisible} and
\emph{indivisible} dynamics. Section~\ref{sec:The-Stochastic-Quantum-Correspondence} will describe
the stochastic-quantum correspondence in detail, including the notion
of a \emph{division event}. Section~\ref{sec:Measurements} will provide a detailed treatment
of the measurement process, which will entail introducing the notion
of an \emph{emergeable}, and then turn to a larger analysis of the
measurement problem and the uncertainty principle.  Section~\ref{sec:Discussion-and-Future-Work}
will conclude the paper with a brief discussion, which will include
identifying a fundamental \emph{category problem} in textbook versions
of quantum theory, as well as describe several open questions to be
addressed in future work.

\section{Indivisible Stochastic Processes\label{sec:Indivisible-Stochastic-Processes}}

\subsection{Basic definitions\label{subsec:Basic-Definitions}}

An \emph{indivisible stochastic process}\footnote{For pedagogical treatments of the theory of stochastic processes,
see the textbooks by Rosenblatt (1962), Parzen (1962), Doob (1990),
or Ross (1995)\nocite{Rosenblatt:1962rp,Parzen:1962sp,Doob:1990sp,Ross:1995sp}.} will be defined as a model consisting of two basic ingredients:
a \emph{configuration space} $\configspace$;  and a dynamical law
in the form of a family of \emph{transition maps} $\stochasticmatrix_{t\from t_{0}}$
that act linearly on probability distributions over $\configspace$
at times $t_{0}$ from some index set, called \emph{conditioning times},
to yield corresponding probability distributions over $\configspace$
at times $t$ from some possibly distinct index set, called \emph{target times}.
The configuration space $\configspace$ (the kinematics) and the transition
maps $\stochasticmatrix_{t\from t_{0}}$ (the dynamics) will constitute
the \emph{fixed} features of the model, whereas the probability distributions
will be \emph{contingent} features allowed to vary from one physical
instantiation or run of the model to another.

For the purposes of this paper, the set of target times $t$ will
usually be assumed to be isomorphic to the real line $\mathbb{R}$,
up to a choice of measurement units. The set of conditioning times
$t_{0}$ will be assumed to contain at least one element, which can
be taken to be the ``initial time'' $0$ without loss of generality.
Note that the target time $t$ is treated here as a real-valued variable
that can be zero, positive, or negative, so there is no assumption
of any fundamental breaking of time-reversal invariance. The choice
of conditioning times might appear to single out the initial time
$0$ as a special time, but Subsection~\ref{subsec:Division-Events-and-the-Markov-Approximation}
will show that for systems in sufficiently strong contact with a repeatedly
eavesdropping environment, as would be the case for generic macroscopic
systems, the initial time $0$ will typically be only one of many
conditioning times that play a similar role.

The formalism for an indivisible stochastic process is easiest to
express in the case in which the system's configuration space $\configspace\defeq\set{1,\dots,N}$
has a finite number\footnote{All the formulas ahead can be extended to systems with continuous
configuration spaces. For ease of exposition, the finite, discrete
case will be assumed going forward. Bear in mind that ``finite''
can be \emph{extremely} large, and ``discrete'' can be well below
any feasible experimental sensitivity or resolution.} $N$ of configurations labeled by positive integers $1,\dots,N$,
perhaps under a suitable form of coarse-graining. In that case, the
system's \emph{standalone probabilities} at a conditioning time $t_{0}$
can be denoted by $p_{j}\left(t_{0}\right)$, the standalone probabilities
at a target time $t$ can be denoted by $p_{i}\left(t\right)$, and
the transition maps $\stochasticmatrix_{t\from t_{0}}$ consist of
\emph{conditional probabilities} 
\begin{equation}
\stochasticmatrix_{ij}\left(t\from t_{0}\right)\defeq p\left(i,t\given j,t_{0}\right),\label{eq:DefStochasticMatrixAsConditionals}
\end{equation}
each of which is the conditional probability for the system to be
in its $i$th configuration at the target time $t$, given that the
system is in its $j$th configuration at the conditioning time $t_{0}$.
Being probabilities, these quantities satisfy the usual non-negativity
conditions 
\begin{equation}
p_{j}\left(t_{0}\right),p_{i}\left(t\right),\stochasticmatrix_{ij}\left(t\from t_{0}\right)\geq0,\label{eq:NonNegativityProbabilities}
\end{equation}
 as well as the normalization conditions 
\begin{equation}
\sum_{j=1}^{N}p_{j}\left(t_{0}\right)=\sum_{i=1}^{N}p_{i}\left(t\right)=\sum_{i=1}^{N}\stochasticmatrix_{ij}\left(t\from t_{0}\right)=1.\label{eq:NormalizationProbabilities}
\end{equation}
 Then from the law of total probability, or marginalization, $p_{i}\left(t\right)=\sum_{j=1}^{N}p\left(i,t\given j,t_{0}\right)p_{j}\left(t_{0}\right)$,
one has the \emph{linear} relationship 
\begin{equation}
p_{i}\left(t\right)=\sum_{j=1}^{N}\stochasticmatrix_{ij}\left(t\from t_{0}\right)p_{j}\left(t_{0}\right),\label{eq:FinalStandaloneProbabilitiesFromMarginalizationStochasticMatrix}
\end{equation}
 where the standalone probabilities $p_{j}\left(t_{0}\right)$ at
the conditioning time $t_{0}$ are assumed to be arbitrary and contingent,
and can therefore be freely adjusted without altering the conditional
probabilities $\stochasticmatrix_{ij}\left(t\from t_{0}\right)$,
which are regarded as fixed features of the model.

Let $p\left(t_{0}\right)$ denote an $N\times1$ \emph{probability vector}
whose entries are given by the standalone probabilities $p_{j}\left(t_{0}\right)$,
$p\left(t\right)$ denote the analogous $N\times1$ probability vector
with entries given by $p_{i}\left(t\right)$, and $\stochasticmatrix\left(t\from t_{0}\right)$
denote the $N\times N$ time-dependent \emph{transition matrix}
consisting of the conditional probabilities $\stochasticmatrix_{ij}\left(t\from t_{0}\right)$.
Then one can naturally recast the linear marginalization relationship
\eqref{eq:FinalStandaloneProbabilitiesFromMarginalizationStochasticMatrix}
in matrix form as 
\begin{equation}
p\left(t\right)=\stochasticmatrix\left(t\from t_{0}\right)p\left(t_{0}\right).\label{eq:MatrixFormStochasticMap}
\end{equation}
 The non-negativity and normalization conditions on the time-dependent
transition matrix $\stochasticmatrix\left(t\from t_{0}\right)$ identify
it as a \emph{(column) stochastic matrix} for each pair of times
$t$ and $t_{0}$. On physical grounds, $\stochasticmatrix\left(t\from t_{0}\right)$
will be assumed to satisfy the continuity condition that in the limit
$t\to t_{0}$, it approaches its value $\stochasticmatrix\left(t_{0}\from t_{0}\right)$,
which will be taken to be the $N\times N$ identity matrix $\idmatrix\defeq\diag{1,\dots,1}$.

Crucially, the transition matrix $\stochasticmatrix\left(t\from t_{0}\right)$
will \emph{not} be assumed to be \textquoteleft divisible,\textquoteright{}
a term that seems to have originated in the research literature in
a 2008 paper by Wolf and Cirac (2008)\nocite{WolfCirac:2008dqc} in
the context of quantum channels.\footnote{Note that this notion of divisibility is conceptually distinct from
the much older concept of \emph{infinite divisibility}, which refers
to a probability distribution that can be expressed as the probability
distribution of a sum of any integer number of independent and identically
distributed random variables.} That is, $\stochasticmatrix\left(t\from t_{0}\right)$ will generically
be \emph{indivisible} (Milz, Modi 2021)\nocite{MilzModi:2021qspaqnp},
meaning that for intermediate times $t^{\prime}$ satisfying $t>t^{\prime}>t_{0}$,
there will not generally exist a genuinely stochastic matrix $\tilde{\stochasticmatrix}\left(t\from t^{\prime}\right)$
satisfying the composition law or \emph{divisibility condition}
\begin{equation}
\stochasticmatrix\left(t\from t_{0}\right)=\tilde{\stochasticmatrix}\left(t\from t^{\prime}\right)\stochasticmatrix\left(t^{\prime}\from t_{0}\right).\label{eq:DivisibilityCondition}
\end{equation}
 In particular, the stochastic process based on the transition matrix
$\stochasticmatrix\left(t\from t_{0}\right)$ will generically fail
to be Markovian, so its dynamical laws will not be iterative over
time in the sense of repeated matrix multiplication $\stochasticmatrix\stochasticmatrix\cdots\stochasticmatrix$,
and the model will also lack specific dynamical laws describing transitions
between arbitrarily chosen intermediate times.

For small configuration spaces, it is easy to devise  smooth, time-dependent,
non-Markovian, indivisible transition matrices. Examples include $2\times2$
transition matrices of the form 
\begin{equation}
\stochasticmatrix\left(t\from0\right)\defeq\begin{pmatrix}f\left(t\right) & 1-f\left(t\right)\\
1-f\left(t\right) & f\left(t\right)
\end{pmatrix}\label{eq:DefSimple2x2StochasticMatrix}
\end{equation}
 for $f\left(t\right)\defeq\exp\left(-t^{2}/\tau^{2}\right)$, with
$\tau$ a constant with units of time, or for $f\left(t\right)\defeq\cos^{2}\omega t$,
with $\omega$ a constant with units of inverse-time. These two time-dependent
transition matrices are provably indivisible, because any matrix $\tilde{\stochasticmatrix}\left(t\from t^{\prime}\right)$
satisfying the divisibility condition above would need to have negative
entries for at least some pairs of times $t$ and $t^{\prime}$, and
would therefore not be a genuine stochastic matrix.

Next, consider a \emph{random variable} $A\left(t\right)$ with (not
necessarily distinct) real-valued \emph{magnitudes}\\
$a_{1}\left(t\right),\dots,a_{N}\left(t\right)$ determined by the
system's configuration $i=1,\dots,N$ and possibly also depending
explicitly on the time $t$.  The \emph{expectation value} $\expectval{A\left(t\right)}$
is then defined as the statistical average of $A\left(t\right)$
over the system's standalone probability distribution at $t$: 
\begin{equation}
\expectval{A\left(t\right)}\defeq\sum_{i=1}^{N}a_{i}\left(t\right)p_{i}\left(t\right).\label{eq:DefRandomVariableExpectationValue}
\end{equation}
 One can define various statistical moments of $A\left(t\right)$
by appropriate generalizations of this basic definition.

\subsection{Markovian and non-Markovian stochastic processes\label{subsec:Markovian-and-Non-Markovian-Stochastic-Processes}}

In general, the dynamical laws of a \emph{non-Markovian stochastic process}
consist of a tower of conditional probabilities of arbitrary order:
\begin{equation}
\begin{aligned} & p\left(i,t\right) &  & \left(\textrm{zeroth order}\right),\\
 & p\left(i,t\given j_{1},t_{1}\right) &  & \left(\textrm{first order}\right),\\
 & p\left(i,t\given j_{1},t_{1};j_{2},t_{2}\right) &  & \left(\textrm{second order}\right),\\
 & p\left(i,t\given j_{1},t_{1};j_{2},t_{2};j_{3},t_{3}\right) &  & \left(\textrm{third order}\right),
\end{aligned}
\label{eq:ConditionalProbabilitiesTower}
\end{equation}
 and so forth. From these conditional probabilities, one can use the
basic rules of probability theory to construct all joint and standalone
probabilities at all choices of times, such as, say, three-time joint
probabilities 
\begin{equation}
p\left(i_{1},t_{1};i_{2},t_{2};i_{3},t_{3}\right)=p\left(i_{1},t_{1}\given i_{2},t_{2};i_{3},t_{3}\right)p\left(i_{2},t_{2}\given i_{3},t_{3}\right)p\left(i_{3},t_{3}\right).\label{eq:ThreeTimeJointProbabilityFromTower}
\end{equation}

Specifying a particular non-Markovian stochastic process uniquely
would therefore require providing an infinite amount of information
in the form of the tower of arbitrary-order conditional probabilities
\eqref{eq:ConditionalProbabilitiesTower}. Moreover, all the joint
probabilities that are definable from this tower of arbitrary-order
conditional probabilities would then need to be related to each other
by an intricate web of marginalization operations, such as 
\begin{equation}
p\left(i_{1},t_{1};i_{3},t_{3}\right)=\sum_{i_{2}}p\left(i_{1},t_{1};i_{2},t_{2};i_{3},t_{3}\right).\label{eq:TwoTimeJointProbabilityFromMarginalization}
\end{equation}

One traditional approach for avoiding these difficulties is to make
the \emph{Markov approximation}, which leads to a \emph{Markov process}
or \emph{Markovian stochastic process}. According to the Markov approximation,
one assumes that any higher-order conditional probability of the form
$p\left(i,t\given j_{1},t_{1};j_{2},t_{2};\dots\right)$ is equal
to the first-order conditional probability $p\left(i,t\given j_{k},t_{k}\right)$
for which $t_{k}$ is the closest conditioning time to the target
time $t$ satisfying $t_{k}<t$. 

An indivisible stochastic process represents an alternative approach
in which one avoids making the Markov approximation but instead works
with \emph{equivalence classes} of non-Markovian processes. In detail,
one considers the entire equivalence class of non-Markovian stochastic
processes that may differ in their higher-order conditional probabilities
but share the same first-order conditional probabilities $p\left(i,t\given j,t_{0}\right)$
defined in \eqref{eq:DefStochasticMatrixAsConditionals}. As will
be shown in the work ahead, these first-order conditional probabilities
will be enough to give agreement with all the empirical predictions
of quantum theory.

To the extent that quantum theory is empirically adequate, the higher-order
conditional probabilities are then unobservable in experiments and
will be left unspecified in this paper. In particular, probabilities
assigned to \emph{whole trajectories}, as constructed from higher-order
conditional probabilities in the sense of \eqref{eq:ThreeTimeJointProbabilityFromTower},
are then left unspecified as well. The higher-order conditional probabilities
of an indivisible stochastic process could, in principle, vary contingently
from one set of instantiations or runs of the process to another.
Whether there exists some theoretical principle that picks out one
set of higher-order or whole-trajectory probabilities from all the
various possibilities is a question that will be left to future work.

\section{The Stochastic-Quantum Correspondence\label{sec:The-Stochastic-Quantum-Correspondence}}

\subsection{The dictionary\label{subsec:The-Dictionary}}

One of the goals of this paper will be to introduce a new and highly
general framework for formulating time-dependent transition matrices
$\stochasticmatrix\left(t\from t_{0}\right)$, conceptually akin to
the Lagrangian or Hamiltonian frameworks for formulating deterministic
dynamics for mechanical systems.

For purposes of notational simplicity, the conditioning time $t_{0}$
will now be taken to be the ``initial time'' $0$. The starting
place will then be to \textquoteleft solve\textquoteright{} the non-negativity
condition $\stochasticmatrix_{ij}\left(t\from0\right)\geq0$ on the
individual entries of the transition matrix $\stochasticmatrix\left(t\from0\right)$
by expressing them in the following way: 
\begin{equation}
\stochasticmatrix_{ij}\left(t\from0\right)=\verts{\dynop_{ij}\left(t\from0\right)}^{2}.\label{eq:StochasticMatrixEntryFromAbsValSquare}
\end{equation}
 Keep in mind that this equation is an entry-by-entry by statement
and does not involve the standard rule for matrix multiplication.\footnote{This expression for the transition matrix $\stochasticmatrix\left(t\from0\right)$
can be regarded as a factorization of the form $\stochasticmatrix\left(t\right)=\overconj{\dynop\left(t\right)}\hadamardprod\dynop\left(t\right),$
where the overbar denotes complex conjugation and where $\hadamardprod$
is the \emph{Schur-Hadamard product} defined for arbitrary $N\times N$
matrices $X$ and $Y$ as entry-wise multiplication: $\left(X\hadamardprod Y\right)_{ij}\defeq X_{ij}Y_{ij}$
(Shur 1911, Horn 1990)\nocite{Schur:1911bztdbbmuvv,Horn:1990thp}.} Note also that this equation is not a postulate\textemdash it is
a mathematical identity.

The $N\times N$ matrix $\dynop\left(t\from0\right)$ introduced here
is guaranteed to exist, although it is not unique, so one can view
it as metaphorically akin to a ``potential'' for $\stochasticmatrix\left(t\from0\right)$,
in analogy with the relationship between a potential energy and a
Newtonian force.\footnote{This nonuniqueness implies a previously unrecognized form of gauge
invariance for all quantum systems, in which one changes the individual
entries $\dynop_{ij}\left(t\from0\right)$ by arbitrary, time-dependent
phase factors: $\dynop_{ij}\left(t\from0\right)\mapsto\exp\left(\theta_{ij}\left(t\right)\right)\dynop_{ij}\left(t\from0\right)$.
These gauge transformations then alter the structure of the resulting
Hilbert-space representation ahead, including the dynamics, in such
a way that all empirical results remain unchanged. Due to space limitations,
nothing more will be discussed in this paper about this novel form
of gauge invariance, which is distinct from an altogether different
form of highly general gauge invariance introduced by Brown (1999)\nocite{Brown:1999aooiqm}.} Its entries $\dynop_{ij}\left(t\from0\right)$ could be taken to
be the real square roots of the corresponding quantities $\stochasticmatrix_{ij}\left(t\from0\right)$,
but they could also include complex numbers, quaternions, or even
the elements of a more general  algebra. With the eventual goal
of reproducing the usual Hilbert-space formalism of quantum theory,
this paper will choose $\dynop_{ij}\left(t\from0\right)$ to involve
only the complex numbers at most.

Due to the normalization condition on the transition matrix $\stochasticmatrix\left(t\from0\right)$,
the matrix $\dynop\left(t\from0\right)$ must satisfy the \emph{summation condition}
\begin{equation}
\sum_{i=1}^{N}\verts{\dynop_{ij}\left(t\from0\right)}^{2}=1.\label{eq:SumTimeEvOpAbsValSqEq1}
\end{equation}
 For now, no further conditions, such as unitarity, will be imposed
on $\dynop\left(t\from0\right)$, whose significance will soon become
more clear.

 Let $e_{1},\dots,e_{N}$ define the system's \emph{configuration basis},
where $e_{i}$ has a $1$ in its $i$th entry and $0$s in all its
other entries. With $\adj$ as the standard adjoint operation, let
\begin{equation}
\projector_{i}\defeq e_{i}e_{i}^{\adj}=\mathrm{diag}\parens{0,\dots,0,\underset{\mathclap{\substack{\uparrow\\
i\textrm{th entry}
}
}}{1},0,\dots,0},\label{eq:DefConfigurationProjectors}
\end{equation}
 denote a \emph{configuration projector}, which is an $N\times N$
matrix consisting of a single $1$ in its $i$th row and  $i$th
column, and $0$s in all its other entries. Letting $\tr\left(\cdots\right)$
denote the usual trace, one can then recast the identity $\stochasticmatrix_{ij}\left(t\from0\right)=\verts{\dynop_{ij}\left(t\from0\right)}^{2}$
relating the entries of $\stochasticmatrix\left(t\from0\right)$ with
the entries of $\dynop\left(t\from0\right)$ as  
\begin{equation}
\keyeq{\stochasticmatrix_{ij}\left(t\from0\right)=\tr\parens{\dynop^{\adj}\left(t\from0\right)\projector_{i}\dynop\left(t\from0\right)\projector_{j}}.}\label{eq:DefDictionary}
\end{equation}
 This equation is a new result and will turn out to serve as the
basic \emph{dictionary} of the \emph{stochastic-quantum correspondence}.
This dictionary translates between the formalism of indivisible stochastic
processes, as symbolized by $\stochasticmatrix_{ij}\left(t\from0\right)$
on the left-hand side, and an expansive set of mathematical tools
for constructing stochastic dynamics, as embodied by the right-hand
side.\footnote{Similar-looking formulas appear incidentally in the equations (3)\textendash (6)
of Auffeves and Gragnier (2017)\nocite{AuffevesGrangier:2017rtqffpra}
as an intermediate step in proving a lemma that the authors use for
conceptually different purposes.}

\subsection{The Hilbert-space representation\label{subsec:The-Hilbert-Space-Representation}}

The matrix $\dynop\left(t\from0\right)$ belongs to the set of operators
acting on a \emph{Hilbert space}, meaning a complete inner-product
space over the complex numbers. More explicitly, $\dynop\left(t\from0\right)$
picks out a Hilbert space $\hilbspace\cong\mathbb{C}^{N}$ that is
isomorphic to the vector space $\mathbb{C}^{N}$ of $N\times1$ column
vectors $v,w,\dots$ with complex-valued entries, under the inner
product $v^{\adj}w$. One therefore arrives at a \emph{Hilbert-space formulation}
for constructing highly generic forms of stochastic dynamics.

The linear marginalization relationship \eqref{eq:FinalStandaloneProbabilitiesFromMarginalizationStochasticMatrix},
$p_{i}\left(t\right)=\sum_{j=1}^{N}\stochasticmatrix_{ij}\left(t\from0\right)p_{j}\left(0\right)$,
between the system's standalone probabilities $p_{j}\left(0\right)$
at the initial time $0$ and the standalone probabilities $p_{i}\left(t\right)$
at the target time $t$ can now be recast as 
\begin{equation}
p_{i}\left(t\right)=\tr\parens{\projector_{i}\densitymatrix\left(t\right)}.\label{eq:FinalStandaloneProbabilitiesFromTraceDensityMatrix}
\end{equation}
 Here 
\begin{equation}
\densitymatrix\left(t\right)\defeq\dynop\left(t\from0\right)\left[\sum_{j=1}^{N}p_{j}\left(0\right)\projector_{j}\right]\dynop^{\adj}\left(t\from0\right)=\dynop\left(t\right)\diag{\dots,p_{j}\left(0\right),\dots}\dynop^{\adj}\left(t\right)\label{eq:DefTimeDependentDensityMatrix}
\end{equation}
 is an $N\times N$ time-dependent matrix that is positive semidefinite,
$\densitymatrix\left(t\right)\geq0$, is self-adjoint, $\densitymatrix^{\adj}\left(t\right)=\densitymatrix\left(t\right)$,
has unit trace, $\tr\parens{\densitymatrix\left(t\right)}=1$, and
is generically non-diagonal. Crucially, notice how the linearity of
the marginalization relationship \eqref{eq:FinalStandaloneProbabilitiesFromMarginalizationStochasticMatrix}
is ultimately responsible for the linearity of the relationship between
the matrix $\densitymatrix\left(t\right)$ and its value $\densitymatrix\left(0\right)$
at the initial time $0$.

Similarly, by substituting $p_{i}\left(t\right)=\tr\parens{\projector_{i}\densitymatrix\left(t\right)}$
from \eqref{eq:FinalStandaloneProbabilitiesFromTraceDensityMatrix}
into the definition $\expectval{A\left(t\right)}\defeq\sum_{i=1}^{N}a_{i}\left(t\right)p_{i}\left(t\right)$
of the expectation value of a random variable $A\left(t\right)$,
one obtains 
\begin{equation}
\expectval{A\left(t\right)}=\tr\parens{A\left(t\right)\densitymatrix\left(t\right)}.\label{eq:ExpectationValueRandomVariableFromTrace}
\end{equation}
 Here $A\left(t\right)$ is now understood to be the $N\times N$
time-dependent, diagonal matrix whose entries are the random variable's
individual magnitudes $a_{1}\left(t\right),\dots,a_{N}\left(t\right)$:
\begin{equation}
A\left(t\right)\defeq\sum_{i=1}^{N}a_{i}\left(t\right)\projector_{i}=\diag{\dots,a_{i}\left(t\right),\dots}.\label{eq:DefDiagonalRandomVariableMatrix}
\end{equation}

In the special case in which the system's standalone probability distribution
at the initial time $0$ is pure, meaning that one of the system's
configurations $j$ is occupied with probability $1$, the system's
probability vector at the initial time $0$ is equal to the $j$th
configuration basis vector $e_{j}$, which again has a $1$ in its
$j$th entry and $0$s in all its other entries. One can then define
an $N\times1$ column vector 
\begin{equation}
\Psi\left(t\right)\defeq\dynop\left(t\from0\right)e_{j},\label{eq:DefStateVector}
\end{equation}
 which is ultimately just the $j$th column of $\dynop\left(t\from0\right)$.
Due to the summation condition $\sum_{i=1}^{N}\verts{\dynop_{ij}\left(t\from0\right)}^{2}=1$
from \eqref{eq:SumTimeEvOpAbsValSqEq1}, this column vector $\Psi\left(t\right)$
automatically has unit norm according to 
\begin{equation}
\sqrt{\Psi^{\adj}\left(t\right)\Psi\left(t\right)}=1.\label{eq:StateVectorUnitNorm}
\end{equation}
 Moreover, the $i$th component $\Psi_{i}\left(t\right)$ of $\Psi\left(t\right)$
is equal to the specific complex-valued matrix entry $\dynop_{ij}\left(t\from0\right)$.
This component $\Psi_{i}\left(t\right)$ is a purely law-like quantity,
in the sense of being just another name for a part of $\dynop\left(t\from0\right)$,
which is itself just a way of encoding the system's dynamical law,
as embodied by the system's transition matrix $\stochasticmatrix\left(t\from0\right)$.

It follows from a short calculation that when the purity condition
$\Psi\left(0\right)=e_{j}$ holds at the initial time $0$, the self-adjoint
matrix $\densitymatrix\left(t\right)$ defined above is rank-one and
has factorization 
\begin{equation}
\densitymatrix\left(t\right)=\Psi\left(t\right)\Psi^{\adj}\left(t\right).\label{eq:RankOneDensityMatrixFactorizedStateVector}
\end{equation}
 The probability formula $p_{i}\left(t\right)=\tr\parens{\projector_{i}\densitymatrix\left(t\right)}$
from \eqref{eq:FinalStandaloneProbabilitiesFromTraceDensityMatrix}
then simplifies to 
\begin{equation}
p_{i}\left(t\right)=\verts{\Psi_{i}\left(t\right)}^{2},\label{eq:ConfigurationBornRuleFromStateVector}
\end{equation}
 and the formula $\expectval{A\left(t\right)}=\tr\parens{A\left(t\right)\densitymatrix\left(t\right)}$
from \eqref{eq:ExpectationValueRandomVariableFromTrace} for the expectation
value of a random variable $A\left(t\right)$ becomes 
\begin{equation}
\expectval{A\left(t\right)}=\Psi^{\adj}\left(t\right)A\left(t\right)\Psi\left(t\right).\label{eq:ExpectationValueRandomVariableFromStateVector}
\end{equation}

Looking at all these results, one notices a striking resemblance to
mathematical objects and formulas that are familiar from textbook
quantum theory.\footnote{For pedagogical treatments of quantum theory, see the textbooks by
Griffiths and Schroeter (2018); Townsend (2012); Shankar (1994); Sakurai
and Napolitano (2010); and Schumacher and Westmoreland (2010)\nocite{GriffithsSchroeter:2018iqm,Townsend:2012maqm,Shankar:1994pqm,SakuraiNapolitano:2010mqm,SchumacherWestmoreland:2010qpsi}.} Specifically, one sees that $\dynop\left(t\from0\right)$ plays the
role of a \emph{time-evolution operator}, $\densitymatrix\left(t\right)$
is a \emph{density matrix,} $\Psi\left(t\right)$ is a \emph{state vector}
or \emph{wave function}, and $A\left(t\right)$ represents an \emph{observable}.\footnote{Note that for the purposes of this paper, the terms \textquoteleft operator\textquoteright{}
and \textquoteleft matrix\textquoteright{} will be used interchangeably,
as will the terms \textquoteleft state vector\textquoteright{} and
\textquoteleft wave function.\textquoteright{}} The probability formulas $p_{i}\left(t\right)=\tr\parens{\projector_{i}\densitymatrix\left(t\right)}$
and $p_{i}\left(t\right)=\verts{\Psi_{i}\left(t\right)}^{2}$ coincide
with the \emph{Born rule}, and $\expectval{A\left(t\right)}=\tr\parens{A\left(t\right)\densitymatrix\left(t\right)}$
and $\expectval{A\left(t\right)}=\Psi^{\adj}\left(t\right)A\left(t\right)\Psi\left(t\right)$
have the same form as the standard expressions for quantum expectation
values.

Despite the similarity to expressions found in quantum theory, as
well as the appearance of non-diagonal matrices, it is important to
keep in mind that the system under investigation here is always fundamentally
in a specific configuration $i=1,\dots,N$ in its configuration space
$\configspace$ at any given time and that the system's dynamics
is completely captured by the transition matrix $\stochasticmatrix\left(t\from0\right)$,
whose entries are conditional probabilities $p\left(i,t\given j,0\right)$.
The mathematical objects $\dynop\left(t\from0\right)$, $\densitymatrix\left(t\right)$,
$\Psi\left(t\right)$, and $A\left(t\right)$, despite being extremely
useful, do not naturally have direct physical meanings, in part because
they are not uniquely defined by $\configspace$ or by $\stochasticmatrix\left(t\from0\right)$.

\subsection{Kraus decompositions\label{subsec:Kraus-Decompositions}}

In the most general case, a time-evolution operator $\dynop\left(t\from0\right)$
may not satisfy any nontrivial constraints apart from the summation
condition $\sum_{i=1}^{N}\verts{\dynop_{ij}\left(t\from0\right)}^{2}=1$
from \eqref{eq:SumTimeEvOpAbsValSqEq1}. It will turn out to be helpful
to find alternative ways of representing the $N\times N$ matrix $\dynop\left(t\from0\right)$
in terms of more tightly constrained mathematical objects.

For $\beta=1,\dots,N$, and with $\projector_{\beta}$ the corresponding
configuration projector defined in \eqref{eq:DefConfigurationProjectors},
let $\krausmatrix_{\beta}\left(t\from0\right)\defeq\dynop\left(t\from0\right)\projector_{\beta}$
be the $N\times N$ matrix defined to share its $\beta$th column
with $\dynop\left(t\from0\right)$, but with $0$s in all its other
entries: 
\begin{equation}
\krausmatrix_{\beta,ij}\left(t\from0\right)\defeq\left(\dynop\left(t\from0\right)\projector_{\beta}\right)_{ij}\defeq\delta_{\beta j}\dynop_{ij}\left(t\from0\right)=\begin{cases}
\dynop_{ij}\left(t\from0\right) & \textrm{for }\beta=j,\\
0 & \textrm{for }\beta\ne j,
\end{cases}\label{eq:DefKrausOperatorEntries}
\end{equation}
 where $\delta_{\beta j}$ is the usual Kronecker delta.  The summation
condition on $\dynop\left(t\from0\right)$ then  becomes the statement
that the matrices $\krausmatrix_{1}\left(t\from0\right),\dots,\krausmatrix_{N}\left(t\from0\right)$
satisfy the \emph{Kraus identity}: 
\begin{equation}
\sum_{\beta=1}^{N}\krausmatrix_{\beta}^{\adj}\left(t\from0\right)\krausmatrix_{\beta}\left(t\from0\right)=\idmatrix.\label{eq:DefKrausIdentity}
\end{equation}
 These matrices are therefore called \emph{Kraus operators} (Kraus
1971)\nocite{Kraus:1971gscqt}. One can then write the basic relationship
$\stochasticmatrix_{ij}\left(t\from0\right)=\verts{\dynop_{ij}\left(t\from0\right)}^{2}$
from \eqref{eq:StochasticMatrixEntryFromAbsValSquare} in an alternative
form called a \emph{Kraus decomposition}: 
\begin{equation}
\stochasticmatrix_{ij}\left(t\from0\right)=\sum_{\beta=1}^{N}\verts{\krausmatrix_{\beta,ij}\left(t\from0\right)}^{2}.\label{eq:StochasticMatrixFromKrausDecomposition}
\end{equation}

Kraus decompositions play a key role in quantum information theory.
They provide (non-unique) generalizations of unitary time evolution
known as \emph{quantum channels}, or \emph{completely positive trace-preserving (CPTP) maps},
that are needed for some kinds of open quantum systems.

\subsection{Unistochastic processes\label{subsec:Unistochastic-Processes}}

The existence of a Kraus decomposition for the time-evolution operator
$\dynop\left(t\from0\right)$ is a crucial new result, and it has
an important corollary. Specifically, if $\dynop\left(t\from0\right)$
is not \emph{already} a unitary matrix, then one can \emph{turn it
into} a unitary matrix by enlarging or \emph{dilating} the original
$N$-element configuration space $\configspace$ to one containing
at most $N^{3}$ configurations. One can then formally regard the
original indivisible stochastic process as a subsystem of this dilated
stochastic process.

In more detail, one starts by combining the given system's $N$-element
configuration space $\configspace$ with an ancillary configuration
space $\configspace^{\prime}$ of some size $N^{\prime}\leq N^{2}$
to yield a dilated configuration space given by the Cartesian product
$\tilde{\configspace}=\configspace\cartesianprod\configspace^{\prime}$,
which then has size $\tilde{N}\leq N^{3}$. By definition, the elements
of this dilated configuration space $\tilde{\configspace}$ take the
form of ordered pairs $\left(i,i^{\prime}\right)$, where $i\in\configspace$
labels the original system's configurations and $i^{\prime}\in\configspace^{\prime}$
labels the configurations of the ancillary system, or \emph{ancilla},
which need not be regarded as physical. The \emph{Stinespring dilation theorem}
(Stinespring 1955, Keyl 2002)\nocite{Stinespring:1955pfoc,Keyl:2002foqit}
then implies the existence of an $\tilde{N}\times\tilde{N}$ \emph{unitary}
time-evolution operator $\tilde{\dynop}\left(t\from0\right)=\tilde{U}\left(t\from0\right)$
whose corresponding $\tilde{N}\times\tilde{N}$ transition matrix
$\tilde{\stochasticmatrix}\left(t\from0\right)$ yields the original
$N\times N$ transition matrix $\stochasticmatrix\left(t\from0\right)$
by marginalization over the ancilla's configuration $i^{\prime}$
at time $t$, for at least some choices of the ancilla's configuration
$j^{\prime}$ at the initial time $0$: $\stochasticmatrix_{ij}\left(t\from0\right)=\sum_{i^{\prime}=1}^{N^{\prime}}\tilde{\stochasticmatrix}_{\left(i,i^{\prime}\right)\left(j,j^{\prime}\right)}\left(t\from0\right)$.
This fact establishes the inevitability of unitary time evolution
in quantum theory. 

Again, the ancilla here need not be treated as a physical subsystem
in its own right. It is important to keep in mind that any empirical
patterns in the observed behavior of the dilated system that become
manifest after formally including the ancilla were already present,
if implicitly, before the formal dilation step.

Without any real loss of generality, the preceding arguments imply
that one can focus on the case in which the time-evolution operator
is unitary, 
\begin{equation}
\dynop\left(t\from0\right)=\timeevop\left(t\from0\right),\label{eq:TimeEvOpUnitary}
\end{equation}
 meaning that 
\begin{equation}
\timeevop^{\adj}\left(t\from0\right)=\timeevop^{-1}\left(t\from0\right).\label{eq:DefUnitaryTimeEvOp}
\end{equation}
 The basic relationship \eqref{eq:StochasticMatrixEntryFromAbsValSquare}
between the system's transition matrix $\stochasticmatrix\left(t\from0\right)$
and the time-evolution operator $\dynop\left(t\from0\right)$ then
becomes 
\begin{equation}
\stochasticmatrix_{ij}\left(t\from0\right)=\verts{\timeevop_{ij}\left(t\from0\right)}^{2}.\label{eq:UnistochasticMatrixFromAbsValSqTimeEvOp}
\end{equation}
 Equivalently, in dictionary form \eqref{eq:DefDictionary}, one has
\begin{equation}
\stochasticmatrix_{ij}\left(t\from0\right)=\tr\parens{\timeevop^{\adj}\left(t\from0\right)\projector_{i}\timeevop\left(t\from0\right)\projector_{j}}.\label{eq:UnistochasticMatrixFromDictionary}
\end{equation}
 The system's transition matrix $\stochasticmatrix\left(t\from0\right)$
is then said to be a \emph{unistochastic matrix}. That is, a unistochastic
matrix is a square matrix whose individual entries are the modulus-squares
of the corresponding entries of a unitary matrix. 

Unistochastic matrices were first introduced in 1954 by Horn (1954)
\nocite{Horn:1954dsmatdoarm}, who originally called them \textquoteleft ortho-stochastic
matrices.\textquoteright{} The modern term \textquoteleft unistochastic
matrix\textquoteright{} was introduced by Thompson in 1989 (Thompson
1989; Nylen, Tam, Uhlig 1993)\nocite{Thompson:1989uln,NylenTamUhlig:1993oteopsonhasm}.
The term \emph{orthostochastic matrix} now refers to a square matrix
whose entries are the modulus-squares of the corresponding entries
of a \emph{real orthogonal} matrix. 

Every orthostochastic matrix is unistochastic. Importantly, however,
the reverse is not generally true, meaning that the complex numbers
generically play a necessary role in formulating a unistochastic transition
matrix $\stochasticmatrix\left(t\from0\right)$ in terms of a unitary
time-evolution operator $\timeevop\left(t\from0\right)$. Even when
the complex numbers are not strictly necessary for writing down a
unitary time-evolution operator $\timeevop\left(t\from0\right)$,
such as if the time-evolution operator can be taken to be real and
orthogonal, it is still very convenient to employ the complex numbers
for a given Hilbert-space representation, so that one can take advantage
of the many useful constructs that show up in standard treatments
of quantum theory, like spectral decompositions and self-adjoint symmetry
generators.\footnote{Intriguingly, time-reversal operators include a complex-conjugation
operator $K$ that \emph{anticommutes} with $i$, meaning that $Ki=-iK$,
so the three mathematical objects $i$, $K$, and $iK$ satisfy $-i^{2}=K^{2}=\left(iK\right)^{2}=iK\left(iK\right)=1$.
They therefore generate a Clifford algebra isomorphic to the \emph{pseudo-quaternions}
(Stueckelberg 1960)\nocite{Stueckelberg:1960qtirhs}. In a sense,
then, the Hilbert spaces of quantum systems are actually defined not
over the complex numbers alone, but over the pseudo-quaternions, although
$K$ is not usually used in the definition of observables.}

It follows immediately from the dictionary formula \eqref{eq:UnistochasticMatrixFromDictionary}
relating $\stochasticmatrix\left(t\from0\right)$ and $\timeevop\left(t\from0\right)$
that every unistochastic transition matrix is \emph{doubly stochastic},
or \emph{bistochastic}, which means that summing over any of its
rows \emph{or} any of its columns always yields $1$: 
\begin{equation}
\sum_{i=1}^{N}\stochasticmatrix_{ij}\left(t\from0\right)=\sum_{j=1}^{N}\stochasticmatrix_{ij}\left(t\from0\right)=1.\label{eq:DoublyStochasticCondition}
\end{equation}

An indivisible stochastic process whose transition matrix $\stochasticmatrix\left(t\right)$
is a unistochastic matrix will be called a \emph{unistochastic process}.

To provide a simple example, note that every permutation matrix $\permutationmatrix$
is, in particular, a unitary matrix. Moreover, because the entries
$\permutationmatrix_{ij}$ of a permutation matrix $\permutationmatrix$
are all $1$s and $0$s, they are individually invariant when one
computes their modulus-squares, so every permutation matrix is \emph{also}
a unistochastic matrix. It follows that a discrete, deterministic
system whose dynamics is defined by a permutation matrix $\permutationmatrix$
is a special case of a unistochastic process.\footnote{Moreover, if $\delta t$ denotes each discrete time step, $n$ denotes
the integer number of time steps, and $t$ denotes a smooth time parameter,
then because real-valued powers of a permutation matrix $\permutationmatrix$
are guaranteed to be unitary, the formula $\stochasticmatrix_{ij}\left(n\,\delta t+t\from n\,\delta t\right)\defeq\verts{\parens{\permutationmatrix^{t/\delta t}}_{ij}}^{2}$
defines a unistochastic matrix that analytically interpolates the
original discrete, deterministic process to a smooth, unistochastic
process.}

Importantly, one can go in the other direction by expressing any unitary
time-evolution operator $\timeevop\left(t\from0\right)$ in terms
of a time-dependent transition matrix $\stochasticmatrix\left(t\from0\right)$
on an underlying configuration space $\configspace$, as noted, for
example, in a paper by Korzekwa and Lostaglio (2021)\nocite{KorzekwaLostaglio:2021qaissp}.
The analysis ahead will explain how to extend this observation into
a new and comprehensive correspondence between indivisible stochastic
processes and quantum systems, going beyond more elementary approaches
that merely embed stochastic processes into a proper subclass of quantum
systems.\footnote{For instance, the \emph{classical-to-classical channels} defined
in the treatment by Wilde (2017)\nocite{Wilde:2017qit} consist of
turning stochastic matrices into a proper subclass of quantum channels
that map diagonal density matrices into diagonal density matrices.}

Assuming a unistochastic process based on a unitary time-evolution
operator $\timeevop\left(t\from0\right)$ that is a differentiable
function of the time $t$, one can define a corresponding self-adjoint
generator $H\left(t\right)=H^{\adj}\left(t\right)$, called the system's
\emph{Hamiltonian}, according to 
\begin{equation}
H\left(t\right)\defeq i\hbar\frac{\partial\timeevop\left(t\from0\right)}{\partial t}\timeevop^{\adj}\left(t\from0\right).\label{eq:DefHamiltonian}
\end{equation}
 Here, the factor of $i$ ensures that the $N\times N$ matrix $H\left(t\right)$
is self-adjoint, and, for present purposes, the \emph{reduced Planck constant}
$\hbar$ is a fixed quantity introduced for purposes of measurement
units. Ultimately, the specific numerical value of $\hbar$ in any
given set of units must be determined empirically by comparison with
experiments.

In terms of the Hamiltonian, the system's density matrix $\densitymatrix\left(t\right)$
then evolves in time according to the \emph{von Neumann equation},
\begin{equation}
i\hbar\frac{\partial\densitymatrix\left(t\right)}{\partial t}=\bracks{H\left(t\right),\densitymatrix\left(t\right)},\label{eq:VonNeumannEq}
\end{equation}
 its state vector $\Psi\left(t\right)$ (if it exists) evolves according
to the \emph{Schrödinger equation}, 
\begin{equation}
i\hbar\frac{\partial\Psi\left(t\right)}{\partial t}=H\left(t\right)\Psi\left(t\right),\label{eq:SchrodingerEq}
\end{equation}
  and its expectation values $\expectval{A\left(t\right)}$ evolve
according to the \emph{Ehrenfest equation}, 
\begin{equation}
\frac{d\angs{A\left(t\right)}}{dt}=\frac{i}{\hbar}\tr\parens{\bracks{H\left(t\right),A\left(t\right)}\densitymatrix\left(t\right)}+\expectval{\frac{\partial A\left(t\right)}{\partial t}}.\label{eq:EhrenfestEquation}
\end{equation}
  Note  that the brackets $\bracks{X,Y}$ that naturally show up
in these equations are genuine \emph{commutators} $XY-YX$, not Poisson
brackets, and involve products of non-diagonal matrices that do not
generally commute with each other under matrix multiplication.

The emergence of these famous equations from a physical model based
on a stochastically evolving trajectory in a configuration space $\configspace$
is a surprising new result.

\subsection{Interference\label{subsec:Interference}}

The appearance of the Schrödinger equation in the previous section
is an important signal that the dictionary \eqref{eq:DefDictionary}
is more than just a tool for using Hilbert-space methods to craft
highly general forms of stochastic dynamics. It also suggests that
indivisible stochastic processes have the resources to replicate the
features of quantum theory more broadly.

As another hint pointing in this direction, one starts by noting again
that an arbitrary time-dependent transition matrix $\stochasticmatrix\left(t\from0\right)$
is generically \emph{indivisible}, in the sense that it does not
satisfy the divisibility condition discussed in  Section~\ref{sec:Indivisible-Stochastic-Processes}
at arbitrary times. To see what goes wrong with divisibility, suppose
that at some time $t^{\prime}$, the transition matrix $\stochasticmatrix\left(t^{\prime}\from0\right)$
has a matrix inverse $\stochasticmatrix^{-1}\left(t^{\prime}\from0\right)$,
and define a new $N\times N$ matrix $\tilde{\stochasticmatrix}\left(t\from t^{\prime}\right)$
according to 
\begin{equation}
\tilde{\stochasticmatrix}\left(t\from t^{\prime}\right)\defeq\stochasticmatrix\left(t\from0\right)\stochasticmatrix^{-1}\left(t^{\prime}\from0\right).\label{eq:DefHypotheticalRelativeStochasticMatrix}
\end{equation}
 As an immediate consequence, one then has 
\begin{equation}
\stochasticmatrix\left(t\from0\right)=\tilde{\stochasticmatrix}\left(t\from t^{\prime}\right)\stochasticmatrix\left(t^{\prime}\from0\right),\label{eq:HypotheticalDivisionEvent}
\end{equation}
 which resembles the divisibility condition \eqref{eq:DivisibilityCondition}.
However, it follows from an elementary theorem of linear algebra that
the inverse of a stochastic matrix can only be stochastic if both
matrices are permutation matrices and, therefore,  do not involve
nontrivial probabilities. Hence, the matrix $\tilde{\stochasticmatrix}\left(t\from t^{\prime}\right)$
defined above is not generically stochastic, so one does not obtain
a genuine form of divisibility.

There is an alternative\textemdash and far-reaching\textemdash way
to understand the generic indivisibility of a time-dependent transition
matrix $\stochasticmatrix\left(t\from0\right)$. To this end, suppose
that $\stochasticmatrix\left(t\from0\right)$ is unistochastic, with
unitary time-evolution operator $\timeevop\left(t\from0\right)$.
Then, for any two times $t$ and $t^{\prime}$, one can define a
\emph{relative} time-evolution operator 
\begin{equation}
\timeevop\left(t\from t^{\prime}\right)\defeq\timeevop\left(t\from0\right)\timeevop^{\adj}\left(t^{\prime}\from0\right),\label{eq:DefRelativeTimeEvOp}
\end{equation}
 which is guaranteed to be unitary and which yields the composition
law 
\begin{equation}
\timeevop\left(t\from0\right)=\timeevop\left(t\from t^{\prime}\right)\timeevop\left(t^{\prime}\from0\right).\label{eq:UnitaryCompositionLaw}
\end{equation}
 Note that this composition law does not extend to the transition
matrix $\stochasticmatrix\left(t\from0\right)$ due to cross terms.

With 
\begin{equation}
\stochasticmatrix_{kj}\left(t^{\prime}\from0\right)\defeq\verts{\timeevop_{kj}\left(t^{\prime}\from0\right)}^{2}\label{eq:UnistochasticInterferenceIntermedTimeAbsValSq}
\end{equation}
 defined as usual, and defining 
\begin{equation}
\stochasticmatrix_{ik}\left(t\from t^{\prime}\right)\defeq\verts{\timeevop_{ik}\left(t\from t^{\prime}\right)}^{2},\label{eq:RelativeUnistochasticInterferenceIntermedTimeAbsValSq}
\end{equation}
 which is manifestly unistochastic, one sees that the discrepancy
between the true transition matrix $\stochasticmatrix\left(t\from0\right)$
and its would-be division $\stochasticmatrix\left(t\from t^{\prime}\right)\stochasticmatrix\left(t^{\prime}\from0\right)$
is given by 
\begin{equation}
\stochasticmatrix_{ij}\left(t\from0\right)-\left[\stochasticmatrix\left(t\from t^{\prime}\right)\stochasticmatrix\left(t^{\prime}\from0\right)\right]_{ij}=\sum_{k\ne l}\overconj{\timeevop_{ik}\left(t\from t^{\prime}\right)\Psi_{k}\left(t^{\prime}\right)}\timeevop_{il}\left(t\from t^{\prime}\right)\Psi_{l}\left(t^{\prime}\right),\label{eq:UnistochasticInterference}
\end{equation}
 where $\Psi\left(t^{\prime}\right)\defeq\timeevop\left(t^{\prime}\from0\right)e_{j}$
is the system's state vector at the time $t^{\prime}$, in keeping
with the general definition of state vectors discussed in Subsection~\ref{subsec:The-Hilbert-Space-Representation},
and where the overbar notation denotes complex conjugation.

Remarkably, the right-hand side of \eqref{eq:UnistochasticInterference}
gives the general mathematical formula for quantum interference, despite
the absence of manifestly quantum-theoretic assumptions. One sees
from this analysis that interference is a direct consequence of the
stochastic dynamics not generally being divisible. More precisely,
interference is nothing more than a generic discrepancy between the
\emph{actual} indivisible stochastic dynamics and a \emph{heuristic-approximate}
divisible stochastic dynamics. Interference encodes the fact that
the underlying stochastic dynamics is indivisible, despite the way
that unitary time-evolution operators look \emph{superficially} divisible.

In particular, quantum-mechanical interference does not imply that
matter has a physically wavelike nature, contrary to frequent claims
in textbook treatments~(Feynman et al. 1965)\nocite{FeynmanLeightonSands:1965tflopv3}.
Indeed, from the perspective of the present discussion, the notion
that quantum-mechanical interference ever necessitated assigning matter
a physically wavelike quality was merely an unfortunate accident of
history, arising from the fact that many early empirical examples
of interference in quantum systems happened to resemble the behavior
of interfering waves propagating in three-dimensional physical space.

These historical examples were clearly special cases. \emph{Multiparticle}
systems have Schrödinger waves that propagate through high-dimensional
configuration spaces, as Schrödinger himself noted in his early work
on wave mechanics (Schrödinger 1926)\nocite{Schrodinger:1926autotmoaam}.
For more abstract systems, like \emph{qubits}, there fail to exist
continuous configuration spaces for Schrödinger waves altogether.

This new way of thinking about quantum-mechanical interference has
implications for the interpretation of the famous \emph{double-slit experiment}.
Recall that in the double-slit experiment, an emitter sends one particle
at a time toward a wall with two slits in it, and a detection screen
on the other side of the wall records the particle's eventual landing
site. In the usual \textquoteleft classical\textquoteright{} description
of the experiment, one asks first which slit the particle enters,
and then, \emph{conditioning} \emph{on the answer}, one then \emph{restarts}
the dynamics with that slit as the new initial condition. Over many
repetitions of the experiment, the detection screen records a statistical
blend from the landing sites of particles passing through the \emph{upper}
slit and particles passing through the \emph{lower} slit. In the case
of quantum-mechanical particles like electrons, however, one instead
finds that the landing sites form a \textquoteleft wavelike\textquoteright{}
interference pattern, and the conclusion is supposedly that each particle
is really a Schrödinger wave of some kind or that the particle fails
to go through one slit or the other.\footnote{The exposition by Feynman, Leighton, and Sands  (1965)\nocite{FeynmanLeightonSands:1965tflopv3}
ends up at precisely such a conclusion: ``It is \emph{not} true that
the electrons go \emph{either} through hole 1 or hole 2.'' {[}Emphasis
in the original.{]} This conclusion, however, does not logically follow
from the empirical appearance of interference effects, but also implicitly
depends on the hidden assumption that the behavior of an electron
in a double-slit experiment can be described by divisible dynamics.}

According to the approach laid out in this paper, the particle always
fundamentally has a single location and is never in both holes simultaneously.
The final interference pattern on the detection screen is not due
to any purported physical reality of Schrödinger waves, but due to
the generic indivisibility of time evolution for quantum systems.
One cannot divide up the particle's evolution into, firstly, its transit
from the emitter to the slits, and then, secondly, conditioned on
which slit the particle enters, the particle's transit from the slits
to the detection screen. The interference that shows up in the double-slit
experiment may be surprising, but that is only because indivisible
stochastic dynamics can be highly unintuitive. In the historical absence
of a sufficiently comprehensive framework for describing indivisible
stochastic dynamics, it was difficult to recognize just how unintuitive
such dynamics could be or what sorts of empirical appearances it
could produce.

In response to this last point, one might suggest that Schrödinger
waves nonetheless offer a superior means of explaining why the double-slit
experiment yields the results that it does. Unfortunately, such hopes
are dashed as soon as one considers sending in \emph{two} particles
on each run of the experiment. A two-particle system's Schrödinger
wave evolves in a \emph{six-dimensional} configuration space, which
is arguably not more physically transparent than indivisible stochastic
dynamics. Indeed, where are the slits supposed to be \emph{located}
in this six-dimensional configuration space?

Of course, if one regards the quantum-mechanical particles that make
up matter as arising more fundamentally from underlying \emph{quantum fields},
then the wavelike properties of those quantum fields ensure that particles
of matter have wavelike properties as well and therefore exhibit
a \emph{wave-particle duality}. That said, there is nothing about
the analysis of the double-slit experiment alone that calls for positing
quantum fields. The necessity of quantum field theory comes from other
theoretical and empirical considerations.\footnote{For a modern motivation, see the textbook by Weinberg (1996)\nocite{Weinberg:1996tqtfi}.}
One should also keep in mind that quantum fields are conceptually
distinct from Schrödinger waves.

\subsection{Implications of interference\label{subsec:Implications-of-Interference}}

The fact that interference shows up in a sufficiently generic stochastic
model means that relative phase factors in state vectors have clear
empirical signatures, even in the absence of the usual axioms of textbook
quantum theory. These empirical manifestations of relative phases
are strong evidence that it should be possible to carry out measurements
on a much wider set of observables than those that are represented
by diagonal matrices in an indivisible stochastic process's configuration
basis. Indeed, Subsection~\ref{subsec:The-Measurement-Process} will
show that non-diagonal, self-adjoint matrices will turn out to be
candidate observables as well.

Thinking more broadly, this overall analysis means that if one is
given an indivisible stochastic process, then there will generically
be a quantitative discrepancy between the system's actual behavior\textemdash as
predicted theoretically or measured empirically\textemdash and predictions
made for the system based on a heuristic-approximate divisible or
Markovian approximation to the system's stochastic dynamics. Again,
this discrepancy is precisely interference.

One way to understand this discrepancy is to note that under a divisibility
approximation, one can assign definite probabilities to each of the
system's possible trajectories by iteratively applying transition
matrices, according to the composition law $\stochasticmatrix\left(t\from0\right)=\stochasticmatrix\left(t\from t^{\prime}\right)\stochasticmatrix\left(t^{\prime}\from0\right)$
from \eqref{eq:DivisibilityCondition}. Iteratively applying transition
matrices is not generically possible for indivisible stochastic processes,
which do not assign unique probabilities to whole trajectories, as
explained in Subsection~\ref{subsec:Markovian-and-Non-Markovian-Stochastic-Processes}.

In the Hilbert-space formulation of an indivisible stochastic process,
one can nonetheless assign complex-valued quantities called \emph{amplitudes}
to the system's possible trajectories, using the fact that unitary
time-evolution operators \emph{can} be composed iteratively, $\timeevop\left(t\from0\right)=\timeevop\left(t\from t^{\prime}\right)\timeevop\left(t^{\prime}\from0\right)$,
as in \eqref{eq:UnitaryCompositionLaw}. These amplitudes form the
conceptual basis for the \emph{path-integral formulation} of quantum
theory (Dirac 1933, Feynman 1942, Feynman 1948)\nocite{Dirac:1933tliqm,Feynman:1942tpolaiqm,Feynman:1948statnrqm}.
From the standpoint of the stochastic-quantum correspondence, which
gives an alternative formulation of quantum theory, the fact that
these amplitudes \textquoteleft interfere\textquoteright{} with each
other does not mean that they all physically occur in some sort of
literal superposition or that the system simultaneously takes all
such paths in reality, but is merely an artifact of the indivisible
dynamics of the underlying indivisible stochastic process.

Collectively, the foregoing observations imply that interference is
not unique to quantum systems but should arise in a much broader
set of physical circumstances. Indeed, given any probabilistically
evolving system with indivisible or non-Markovian dynamics, one should
now be able to interpret any discrepancies between the behavior of
such a system and the behavior of a heuristic-approximate divisible
or Markovian approximation as manifestations of interference. As a
concrete prediction, one could therefore imagine experimentally measuring
interference effects for essentially any system that can be modeled
using indivisible or non-Markovian stochastic dynamics.

\subsection{Division events and the Markov approximation\label{subsec:Division-Events-and-the-Markov-Approximation}}

Why do discrete-time Markov chains provide such a good approximation
to so many stochastic processes in the real world? One intuitively
reasonable explanation is that when a system is not isolated from
a noisy and intrusive environment, delicate correlations from one
time to another \textquoteleft wash out\textquoteright{} over short
time scales as those correlations leak out into the environment.

Deriving this intuitive picture from first principles in a more precise
way might appear to be a difficult task. Indeed, such a derivation
would seem to require finding a more general framework for describing
a non-Markovian process and then showing that such a process becomes
approximately Markovian in the appropriate physical circumstances.
Fortunately, this paper provides just such a framework.

To set things up, one starts by introducing a composite system $\mathcal{S}\mathcal{E}$
consisting of a \emph{subject system} $\mathcal{S}$ together with
an \emph{environment} $\mathcal{E}$. The configurations of the subject
system's configuration space $\configspace_{\mathcal{S}}$ will be
labeled by $i=1,\dots,N$, and the configurations of the environment's
configuration space $\configspace_{\mathcal{E}}$ will be labeled
by $e=1,\dots,M$, where $M\geq N$. The configuration space of the
composite system is then the Cartesian product\footnote{The right-hand side of this equation is indeed a Cartesian product,
not a tensor product, because this equation is a statement about the
composite system's configuration space, not about its dynamics or
Hilbert-space representation.} $\configspace_{\mathcal{S}\mathcal{E}}=\configspace_{\mathcal{S}}\cartesianprod\configspace_{\mathcal{E}}$,
meaning that each element of $\configspace_{\mathcal{S}\mathcal{E}}$
is a simple ordered pair of the form $\left(i,e\right)$. One then
singles out $N$ configurations of the environment by labeling them
as $e\left(1\right),\dots,e\left(N\right)$.

For the dynamics, suppose for simplicity that the composite system
evolves according to an overall unistochastic transition matrix with
individual entries 
\begin{equation}
\stochasticmatrix_{ie,i_{0}e_{0}}^{\mathcal{S}\mathcal{E}}\left(t\from0\right)=\verts{\timeevop_{ie,i_{0}e_{0}}^{\mathcal{S}\mathcal{E}}\left(t\from0\right)}^{2}.\label{eq:CompositeSubjectEnvironmentUnistochasticMatrixFromAbsValSquareUnitary}
\end{equation}
 Furthermore, suppose that the subject system and the environment
interact up to a time $t^{\prime}>0$ in such a way that they end
up with joint probabilities of the form 
\begin{equation}
p_{i^{\prime}e^{\prime}}^{\mathcal{S}\mathcal{E}}\left(t^{\prime}\right)=p_{i^{\prime}}^{\mathcal{S}}\left(t^{\prime}\right)\delta_{e^{\prime}e\parens{i^{\prime}}}.\label{eq:CompositeSubjectEnvironmentIntermedJointProbabilitiesCorrelated}
\end{equation}
 (Note the appearance of a Kronecker delta here.) This formula describes
an idealized \emph{statistical correlation} between the configuration
$i^{\prime}$ of the subject system at $t^{\prime}$ and the corresponding
configuration $e\parens{i^{\prime}}$ of the environment. 

If there is to be any possibility of the two subsystems evolving independently
for times $t>t^{\prime}$ after the interaction has concluded, then
it should be possible to factorize the composite system's relative
time-evolution operator $\timeevop^{\mathcal{S}\mathcal{E}}\left(t\from t^{\prime}\right)$
between the two subsystems for $t>t^{\prime}$ as the following tensor
product:\footnote{Note the natural appearance of a tensor product here, $\timeevop^{\mathcal{S}\mathcal{E}}\left(t\from t^{\prime}\right)=\timeevop^{\mathcal{S}}\left(t\from t^{\prime}\right)\tensorprod\timeevop^{\mathcal{E}}\left(t\from t^{\prime}\right)$,
because this statement refers to the composite system's dynamics in
the system's Hilbert-space representation.} 
\begin{equation}
\timeevop_{ie,i^{\prime}e^{\prime}}^{\mathcal{S}\mathcal{E}}\left(t\from t^{\prime}\right)=\timeevop_{ii^{\prime}}^{\mathcal{S}}\left(t\from t^{\prime}\right)\timeevop_{ee^{\prime}}^{\mathcal{E}}\left(t\from t^{\prime}\right)\quad\textrm{for }t>t^{\prime}.\label{eq:CompositeSubjectEnvironmentTimeEvOpFactorizesAfterInteraction}
\end{equation}

In light of the Born rule, as derived in Subsection~\ref{subsec:The-Hilbert-Space-Representation},
the joint probabilities $p_{i^{\prime}e^{\prime}}^{\mathcal{S}\mathcal{E}}\left(t^{\prime}\right)$
correspond to a wave function\footnote{If necessary, one can easily write down idealized examples of unitary
time-evolution operators for the composite system that produce this
wave function. For instance, one could use $\timeevop^{\mathcal{S}\mathcal{E}}\left(t^{\prime}\from0\right)\defeq\sum_{i^{\prime}}\projector_{i^{\prime}}^{\mathcal{S}}\tensorprod R_{e\parens{i^{\prime}}}^{\mathcal{E}}$,
where $\projector_{i^{\prime}}^{\mathcal{S}}$ is the $i^{\prime}$th
configuration projector for the subject system and where $R_{e\parens{i^{\prime}}}^{\mathcal{E}}$
is a unitary transformation that takes the environment's initial configuration
to the configuration $e\parens{i^{\prime}}$.} 
\begin{equation}
\Psi_{i^{\prime}e^{\prime}}^{\mathcal{S}\mathcal{E}}\left(t^{\prime}\right)=\Psi_{i^{\prime}}^{\mathcal{S}}\left(t^{\prime}\right)\delta_{e^{\prime}e\parens{i^{\prime}}}.\label{eq:CompositeSubjectEnvironmentIntermedWaveFunctionCorrelated}
\end{equation}
 The composite system's wave function at later times $t>t^{\prime}$
after the interaction is therefore given in terms of the relative
time-evolution operator $\timeevop^{\mathcal{S}\mathcal{E}}\left(t\from t^{\prime}\right)$
according to 

\begin{equation}
\Psi_{ie}^{\mathcal{S}\mathcal{E}}\left(t\right)=\sum_{i^{\prime},e^{\prime}}\timeevop_{ie,i^{\prime}e^{\prime}}^{\mathcal{S}\mathcal{E}}\left(t\from t^{\prime}\right)\Psi_{i^{\prime}e^{\prime}}^{\mathcal{S}\mathcal{E}}\left(t^{\prime}\right)=\sum_{i^{\prime}}\timeevop_{ii^{\prime}}^{\mathcal{S}}\left(t\from t^{\prime}\right)\Psi_{i^{\prime}}^{\mathcal{S}}\left(t^{\prime}\right)\timeevop_{ee\parens{i^{\prime}}}^{\mathcal{E}}\left(t\from t^{\prime}\right).\label{eq:CompositeSubjectEnvironmentFinalWaveFunctionCalculation}
\end{equation}
 From the Born rule \eqref{eq:ConfigurationBornRuleFromStateVector},
one sees that the joint probabilities for $t>t^{\prime}$ are given
by 
\begin{equation}
p_{ie}^{\mathcal{S}\mathcal{E}}\left(t\right)=\verts{\Psi_{ie}^{\mathcal{S}\mathcal{E}}\left(t\right)}^{2}.\label{eq:CompositeSubjectEnvironmentFinalJointProbabilitiesFromAbsValSq}
\end{equation}
 Carrying out an ordinary marginalization over the configuration $e$
of the environment and invoking the unitarity of the environment's
relative time-evolution operator $\timeevop^{\mathcal{E}}\left(t\from t^{\prime}\right)$,
one obtains the standalone probabilities $p_{i}^{\mathcal{S}}\left(t\right)$
for the subject system alone for $t>t^{\prime}$: 
\begin{equation}
\begin{aligned} & p_{i}^{\mathcal{S}}\left(t\right)=\sum_{e}p_{ie}^{\mathcal{S}\mathcal{E}}\left(t\right)=\sum_{i_{1}^{\prime},i_{2}^{\prime}}\overconj{\timeevop_{ii_{1}^{\prime}}^{\mathcal{S}}\left(t\from t^{\prime}\right)\Psi_{i_{1}^{\prime}}^{\mathcal{S}}\left(t^{\prime}\right)}\timeevop_{ii_{2}^{\prime}}^{\mathcal{S}}\left(t\from t^{\prime}\right)\Psi_{i_{2}^{\prime}}^{\mathcal{S}}\left(t^{\prime}\right)\sum_{e}\overconj{\timeevop_{ee\parens{i_{1}^{\prime}}}^{\mathcal{E}}\left(t\from t^{\prime}\right)}\timeevop_{ee\parens{i_{2}^{\prime}}}^{\mathcal{E}}\left(t\from t^{\prime}\right)\\
 & =\sum_{i^{\prime}}\verts{\timeevop_{ii^{\prime}}^{\mathcal{S}}\left(t\from t^{\prime}\right)}^{2}\verts{\Psi_{i^{\prime}}^{\mathcal{S}}\left(t^{\prime}\right)}^{2},
\end{aligned}
\label{eq:SubjectFinalStandaloneProbabilitiesCalculation}
\end{equation}
 where, again, the overbar notation denotes complex conjugation.

Taking the limit $t\to t^{\prime}$ and referring back to the Born
rule again, one sees that the subject system's standalone probabilities
at the time $t^{\prime}$ are 
\begin{equation}
p_{i^{\prime}}^{\mathcal{S}}\left(t^{\prime}\right)=\verts{\Psi_{i^{\prime}}^{\mathcal{S}}\left(t^{\prime}\right)}^{2}.\label{eq:SubjectIntermedStandaloneProbabilitiesFromBornRule}
\end{equation}
 One also sees from the last line of the calculation above that, as
in Subsection~\ref{subsec:Interference}, one can identify 
\begin{equation}
\stochasticmatrix_{ii^{\prime}}^{\mathcal{S}}\left(t\from t^{\prime}\right)\defeq\verts{\timeevop_{ii^{\prime}}^{\mathcal{S}}\left(t\from t^{\prime}\right)}^{2}.\label{eq:SubjectRelativeUnistochasticMatrix}
\end{equation}
 Hence, one ends up with a genuinely linear relationship that precisely
mirrors the linear marginalization formula \eqref{eq:FinalStandaloneProbabilitiesFromMarginalizationStochasticMatrix}
introduced in Section~\ref{sec:Indivisible-Stochastic-Processes},
with $t^{\prime}$ now effectively serving as a new \textquoteleft initial
time\textquoteright : 
\begin{equation}
p_{i}^{\mathcal{S}}\left(t\right)=\sum_{i^{\prime}}\stochasticmatrix_{ii^{\prime}}^{\mathcal{S}}\left(t\from t^{\prime}\right)p_{i^{\prime}}^{\mathcal{S}}\left(t^{\prime}\right).\label{eq:SubjectFinalStandaloneProbabilitiesFromMarginalizationRelativeStochasticMatrix}
\end{equation}

Applying the original linear marginalization formula \eqref{eq:FinalStandaloneProbabilitiesFromMarginalizationStochasticMatrix}
from the actual initial time $0$ to the time $t^{\prime}$, one also
has the equation 
\begin{equation}
p_{i^{\prime}}^{\mathcal{S}}\left(t^{\prime}\right)=\sum_{j}\stochasticmatrix_{i^{\prime}j}^{\mathcal{S}}\left(t^{\prime}\from0\right)p_{j}^{\mathcal{S}}\left(0\right).\label{eq:SubjectIntermedStandaloneProbabilitiesFromMarginalizationStochasticMatrix}
\end{equation}
 Combining these results immediately yields 
\begin{equation}
p_{i}^{\mathcal{S}}\left(t\right)=\sum_{j}\stochasticmatrix_{ij}^{\mathcal{S}}\left(t\from0\right)p_{j}^{\mathcal{S}}\left(0\right),\label{eq:SubjectFinalStandaloneProbabilitiesFromMarginalizationStochasticMatrix}
\end{equation}
 where $\stochasticmatrix^{\mathcal{S}}\left(t\from0\right)$ is a
transition matrix that is manifestly \emph{divisible} at $t^{\prime}$:
\begin{equation}
\stochasticmatrix^{\mathcal{S}}\left(t\from0\right)\defeq\stochasticmatrix^{\mathcal{S}}\left(t\from t^{\prime}\right)\stochasticmatrix^{\mathcal{S}}\left(t^{\prime}\from0\right).\label{eq:SubjectDivisibleStochasticMatrixDivisionEvent}
\end{equation}
 Thus, the interaction between the subject system $\mathcal{S}$ and
the environment $\mathcal{E}$ up to the time $t^{\prime}$ has led
to a transition matrix $\stochasticmatrix^{\mathcal{S}}\left(t\from0\right)$
for the subject system that is divisible at $t^{\prime}$, which
has become a valid conditioning time.

It is therefore natural to refer to the new conditioning time $t^{\prime}$
as a \emph{division event}. An important corollary is that the initial
time $0$ is not a unique time but is instead only one of many division
events inevitably experienced by a system in sufficiently strong contact
with a repeatedly eavesdropping environment, in the sense that the
interactions with the environment lead to correlations that look approximately
like those in the formula \eqref{eq:CompositeSubjectEnvironmentIntermedJointProbabilitiesCorrelated}
for $p_{i^{\prime}e^{\prime}}^{\mathcal{S}\mathcal{E}}\left(t^{\prime}\right)$
above.\footnote{Although generically always approximate, division events will become
nearly exact when the environment is sufficiently macroscopic, for
precisely the same reasons that decoherence becomes nearly exact in
such cases. Any resulting discrepancies in the effective stochastic
laws will therefore be minuscule in real-world cases. These tiny discrepancies
in the effective laws for subsystems are inevitable in all no-collapse
formulations or interpretations of quantum theory.} Division events will play a crucial role going forward.

Suppose that these kinds of division events can be approximated as
occurring regularly over a characteristic time scale $\delta t$.
Suppose, moreover, that the unistochastic dynamics is homogeneous
in time, in the sense that $\timeevop^{\mathcal{S}}\left(t+\delta t\from t\right)=\timeevop^{\mathcal{S}}\left(\delta t\from0\right)$
for all times $t$. Then the subject system's transition matrix after
any integer number $n\geq1$ of time steps $\delta t$ is given by
\begin{equation}
\stochasticmatrix^{\mathcal{S}}\left(n\,\delta t\from0\right)=\parens{\stochasticmatrix^{\mathcal{S}}}^{n},\label{eq:SubjectDiscreteMarkovChain}
\end{equation}
 where 
\begin{equation}
\stochasticmatrix_{ij}^{\mathcal{S}}\defeq\verts{\timeevop_{ij}^{\mathcal{S}}\left(\delta t\from0\right)}^{2}.\label{eq:SubjectDiscreteMarkovChainUnistochasticMatrix}
\end{equation}
 The stochastic dynamics therefore takes the form of a discrete-time
Markov chain. This analysis  provides a theoretical explanation for
the ubiquity of Markovian stochastic dynamics in so many real-world
cases, and represents another new result.

In a sense, division events represent a kind of spontaneous breaking
of time-translation symmetry due to interactions between a given
system and its environment. Like other forms of spontaneous symmetry
breaking, division events therefore transcend the line between what
are fixed aspects of the laws and what are contingencies.\footnote{The author would like to thank an anonymous reviewer for requesting
clarification on this point.} 

\subsection{Decoherence\label{subsec:Decoherence}}

Had the environment not interacted with the subject system, then the
subject system's density matrix $\densitymatrix^{\mathcal{S}}\left(t^{\prime}\right)$
at the time $t^{\prime}$ would have generically been non-diagonal,
in accordance with the general definition \eqref{eq:DefTimeDependentDensityMatrix}
provided in  Subsection~\ref{subsec:The-Hilbert-Space-Representation}:
\begin{equation}
\densitymatrix^{\mathcal{S}}\left(t^{\prime}\right)=\timeevop^{\mathcal{S}}\left(t^{\prime}\from0\right)\left[\sum_{j}p_{j}\left(0\right)\projector_{j}\right]\timeevop^{\mathcal{S}\adj}\left(t^{\prime}\from0\right).\label{eq:SubjectIntermedDensityMatrixWithoutDecoherence}
\end{equation}

By contrast, suppose that the environment indeed interacts with the
subject system to produce a division event at $t^{\prime}$, as defined
in the previous subsection. In that case, the standalone probability
$p_{i}^{\mathcal{S}}\left(t\right)$ for the subject system to occupy
its $i$th configuration at $t>t^{\prime}$ is given by the linear
marginalization relationship \eqref{eq:SubjectFinalStandaloneProbabilitiesFromMarginalizationRelativeStochasticMatrix},
which can be written instead as 
\begin{equation}
p_{i}^{\mathcal{S}}\left(t\right)=\tr\parens{\projector_{i}\densitymatrix^{\mathcal{S}}\left(t\right)},\label{eq:SubjectFinalStandaloneProbabilityFromTrace}
\end{equation}
 where 
\begin{equation}
\densitymatrix^{\mathcal{S}}\left(t\right)\defeq\timeevop^{\mathcal{S}}\left(t\from t^{\prime}\right)\densitymatrix^{\mathcal{S}}\left(t^{\prime}\right)\timeevop^{\mathcal{S}\adj}\left(t\from t^{\prime}\right),\label{eq:SubjectFindDensityMatrixFromIntermedWithDecoherence}
\end{equation}
 and where, in turn, 
\begin{equation}
\densitymatrix^{\mathcal{S}}\left(t^{\prime}\right)\defeq\sum_{i^{\prime}}p_{i^{\prime}}^{\mathcal{S}}\left(t^{\prime}\right)\projector_{i^{\prime}}^{\mathcal{S}}=\mathrm{diag}\parens{\dots,p_{i^{\prime}}^{\mathcal{S}}\left(t^{\prime}\right),\dots},\label{eq:SubjectIntermedDensityMatrixWithDecoherence}
\end{equation}
 which is diagonal.

On comparing the two expressions above for the subject system's density
matrix $\densitymatrix\left(t^{\prime}\right)$ at $t^{\prime}$,
one sees that the interaction with the environment has effectively
eliminated the off-diagonal entries, or \emph{coherences}, in the
subject system's density matrix. This phenomenon is called \emph{decoherence},
and the foregoing analysis makes clear that decoherence is nothing
more than the mundane  leakage of correlations into the environment
when viewed through the lens of the Hilbert-space formulation.

This analysis also sheds new light on the meaning of coherences in
density matrices, as well as on \emph{superpositions} in state vectors,
where superpositions are related to coherences in the case of a rank-one
density matrix through the formula $\densitymatrix_{i_{1}i_{2}}\left(t\right)=\Psi_{i_{1}}\left(t\right)\overconj{\Psi_{i_{2}}\left(t\right)}$,
in accordance with the relationship \eqref{eq:RankOneDensityMatrixFactorizedStateVector}
between state vectors and density matrices discussed in  Subsection~\ref{subsec:The-Hilbert-Space-Representation}.
From the standpoint of this analysis, superpositions and coherences
are merely indications that one is catching a given system when it
is in the midst of an indivisible stochastic process, between division
events, rather than implying that the system is literally in \textquoteleft multiple
states at once.\textquoteright{} In other words, coherences and superpositions
are mathematical artifacts of the fundamental indivisibility of the
underlying stochastic process, when represented using a \emph{superficially}
divisible unitary time-evolution operator.

These results may also help explain why the precise connection between
quantum theory and stochastic processes has historically remained
unclear for so long. If one assumes a Markov approximation, as is
often the case in the research literature on stochastic processes,
then coherences and superposition do not show up, meaning that density
matrices remain diagonal, state vectors remain trivial, and nontrivial
unistochastic dynamics cannot arise.

\subsection{Entanglement\label{subsec:Entanglement}}

Consider next a composite system $\mathcal{A}\mathcal{B}$ consisting
of a pair of subsystems $\mathcal{A}$ and $\mathcal{B}$. Suppose
that the two subsystems do not interact from the initial time $0$
up to some later time $t^{\prime}>0$, but then begin interacting
at $t^{\prime}$. 

For times $t$ between $0$ and $t^{\prime}$, the absence of interactions
means that the composite system's transition matrix $\stochasticmatrix^{\mathcal{A}\mathcal{B}}\left(t\from0\right)$
factorizes into the tensor product of a transition matrix $\stochasticmatrix^{\mathcal{A}}\left(t\from0\right)$
for $\mathcal{A}$ and a separate transition matrix $\stochasticmatrix^{\mathcal{B}}\left(t\from0\right)$
for $\mathcal{B}$: 
\begin{equation}
\stochasticmatrix^{\mathcal{A}\mathcal{B}}\left(t\from0\right)=\stochasticmatrix^{\mathcal{A}}\left(t\from0\right)\tensorprod\stochasticmatrix^{\mathcal{B}}\left(t\from0\right)\quad\textrm{for }0\leq t<t^{\prime}.\label{eq:SubsystemsFactorizableStochasticMatrixBeforeEntanglement}
\end{equation}
 Starting at the time $t^{\prime}$, however, the composite system's
transition matrix $\stochasticmatrix^{\mathcal{A}\mathcal{B}}\left(t\from0\right)$,
which encodes cumulative statistical information and therefore correlations,
will fail to tensor-factorize between the two subsystems, in the sense
that 
\begin{equation}
\stochasticmatrix^{\mathcal{A}\mathcal{B}}\left(t\from0\right)\ne\stochasticmatrix^{\mathcal{A}}\left(t\from0\right)\tensorprod\stochasticmatrix^{\mathcal{B}}\left(t\from0\right)\quad\textrm{for }t>t^{\prime}\label{eq:SubsystemsNotFactorizableStochasticMatrixAfterEntanglement}
\end{equation}
 for any possible transition matrices $\stochasticmatrix^{\mathcal{A}}\left(t\from0\right)$
and $\stochasticmatrix^{\mathcal{B}}\left(t\from0\right)$ that properly
capture the respective dynamics of the two subsystems. (It is worth
noting that this loss of tensor-factorization gives a highly general,
model-independent way to \emph{define} an interaction.) Even if the
two subsystems have a notion of localizability in space and are eventually
placed at a large separation distance at some time $t>t^{\prime}$,
the composite system's transition matrix will still fail to tensor-factorize
between the two subsystems, thereby leading to the appearance of what
looks like nonlocal stochastic dynamics across that separation distance.

However, if the composite system exhibits a division event at some
later time $t^{\prime\prime}>t^{\prime}$, perhaps due to interactions
between one of the subsystems and the larger environment, as spelled
out in Subsection~\ref{subsec:Division-Events-and-the-Markov-Approximation},
then the composite system's transition matrix $\stochasticmatrix^{\mathcal{A}\mathcal{B}}\left(t\from0\right)$
will divide at $t^{\prime\prime}$: 
\begin{equation}
\stochasticmatrix^{\mathcal{A}\mathcal{B}}\left(t\from0\right)=\stochasticmatrix^{\mathcal{A}\mathcal{B}}\left(t\from t^{\prime\prime}\right)\stochasticmatrix^{\mathcal{A}\mathcal{B}}\left(t^{\prime\prime}\from0\right)\quad\textrm{for }t>t^{\prime\prime}>t^{\prime}.\label{eq:SubsystemsDivisionEvent}
\end{equation}
 If the two subsystems $\mathcal{A}$ and $\mathcal{B}$ do not interact
with each other after $t^{\prime}$, then the \emph{relative} transition
matrix $\stochasticmatrix^{\mathcal{A}\mathcal{B}}\left(t\from t^{\prime\prime}\right)$
appearing here will tensor-factorize between them, 
\begin{equation}
\stochasticmatrix^{\mathcal{A}\mathcal{B}}\left(t\from t^{\prime\prime}\right)=\stochasticmatrix^{\mathcal{A}}\left(t\from t^{\prime\prime}\right)\tensorprod\stochasticmatrix^{\mathcal{B}}\left(t\from t^{\prime\prime}\right),\label{eq:SubsystemsFactorizableRelativeStochasticMatrixAfterDivisionEvent}
\end{equation}
 so the two subsystems will cease exhibiting what had looked like
nonlocal stochastic dynamics.

This analysis precisely captures the quantum-theoretic notion of
\emph{entanglement}, without any invocation of a Hilbert-space picture.
Systems that interact with each other start to exhibit what superficially
appears to be a nonlocal kind of stochastic dynamics, even if the
systems are moved far apart in physical space, and decoherence by
the environment effectively causes a breakdown in that apparent dynamical
nonlocality.

Due to the stochastic and non-Markovian nature of the laws in this
new formulation of quantum theory, the precise nature of this apparent
dynamical nonlocality is an extremely subtle matter. Relevant questions
concerning locality and causation will be treated in detail in future
work.

\section{Measurements\label{sec:Measurements}}

\subsection{Emergeables\label{subsec:Emergeables}}

The preceding sections have shown that an indivisible stochastic process\textemdash that
is, a physical model with kinematics based on a configuration space
and dynamics based on a suitably non-Markovian stochastic law\textemdash is
capable of accounting for signature features of quantum theory like
superposition, interference, decoherence, and entanglement. In addition,
the Hilbert-space side of the dictionary \eqref{eq:DefDictionary}
contains many expressions and equations that are identical to those
found in quantum theory. 

However, an actual quantum system also includes observables beyond
the random variables introduced in Section~\ref{sec:Indivisible-Stochastic-Processes}\textemdash that
is, beyond the narrow class of observables that are represented by
diagonal matrices. Indeed, the existence of noncommuting observables
represented by self-adjoint matrices that are non-diagonal is another
hallmark feature of quantum theory.

Remarkably, an indivisible stochastic process will generically contain
such observables as well. Specifically, the next subsection will establish
that non-diagonal, self-adjoint matrices represent candidate observables
that naturally satisfy the usual probabilistic rules of quantum theory,
including the Born rule, all without the need to introduce any new
fundamental axioms. In so doing, the analysis ahead will demonstrate
that the dictionary \eqref{eq:DefDictionary} is not merely a tool
for studying a broad class of stochastic processes, but defines a
comprehensive stochastic-quantum correspondence.

These non-diagonal observables therefore resemble random variables
in some ways but represent emergent patterns in the overall stochastic
dynamics for measurement processes and do not have a transparent
interpretation at the level of the underlying configuration space
$\configspace$. These observables will therefore be called \emph{emergeables}.
This terminology is intended for contrast with the system's genuine
random variables, which could be called \emph{beables}\textemdash that
is, \textquoteleft be-ables\textquoteright\textemdash to invoke a
term coined by Bell (1973)\nocite{Bell:1973sao} to refer to observables
that express how a system can physically \emph{be}, ontologically
speaking.

There is a sense in which emergeables are not an entirely new idea,
but are similar to emergent physical properties like temperatures
or pressures that likewise do not have a clear meaning at the level
of a system's fine-grained states. A somewhat more closely related
notion appears in Niels Bohr's famous reply (Bohr 1935, Bell 1971)\nocite{Bohr:1935cqmdoprbcc,Bell:1971itthvq_2}
to the \emph{Einstein-Podolsky-Rosen paradox} (Einstein, Podolsky,
Rosen 1935)\nocite{EinsteinPodolskyRosen:1935cqmdprbcc}, in which
Bohr describes emergent observables that show up in measurement interactions.
These sorts of emergent observables also play a key role in the \emph{de Broglie-Bohm formulation},
or \emph{Bohmian mechanics} (Bell 1982, Daumer et al. 1996)\nocite{Bell:1982otipw,DaumerDurrGoldsteinZanghi:1996nrao},
in which they account for observables other than particle positions.

\subsection{The measurement process\label{subsec:The-Measurement-Process}}

With all the requisite conceptual background now in place, one can
proceed to model the measurement of a generic observable as a physical
process. To start, consider a composite system $\mathcal{S}\mathcal{D}\mathcal{E}$
consisting of three subsystems that will be called a \emph{subject system}
$\mathcal{S}$, a \emph{measuring device} $\mathcal{D}$, and an
\emph{environment} $\mathcal{E}$. Note that one of the additional
goals ahead will be to identify the criteria for a subsystem like
$\mathcal{D}$ to be regarded as a genuine measuring device in the
first place.

Focusing momentarily on the subject system $\mathcal{S}$, consider
an $N\times N$ self-adjoint matrix 
\begin{equation}
\tilde{A}^{\mathcal{S}}=\tilde{A}^{\mathcal{S}\adj},\label{eq:DefSubjectSystemObservableAsSelfAdjoint}
\end{equation}
 which may or may not be one of the subject system's diagonal random
variables.\footnote{More generally, one could take $\tilde{A}^{\mathcal{S}}$ to be a
\emph{normal matrix}, meaning a matrix that commutes with its adjoint
$\tilde{A}^{\mathcal{S}\adj}$.} For example, $\tilde{A}^{\mathcal{S}}$ could be an emergeable like
those introduced in the previous subsection.

By the \emph{spectral theorem}, $\tilde{A}^{\mathcal{S}}$ has a
\emph{spectral decomposition} of the form 
\begin{equation}
\tilde{A}^{\mathcal{S}}=\sum_{\alpha}\tilde{a}_{\alpha}\tilde{\projector}_{\alpha}^{\mathcal{S}},\label{eq:ObservableSpectralDecomposition}
\end{equation}
 where $\tilde{a}_{\alpha}$ are the eigenvalues of $\tilde{A}^{\mathcal{S}}$
and where $\tilde{\projector}_{\alpha}^{\mathcal{S}}$ are its eigenprojectors.
These eigenprojectors $\tilde{\projector}_{\alpha}^{\mathcal{S}}$
are not generically diagonal, but they nonetheless satisfy the mutual
exclusivity condition 
\begin{equation}
\tilde{\projector}_{\alpha}^{\mathcal{S}}\tilde{\projector}_{\alpha^{\prime}}^{\mathcal{S}}=\delta_{\alpha\alpha^{\prime}}\tilde{\projector}_{\alpha}^{\mathcal{S}}\label{eq:ObservableProjectorsMutuallyExclusive}
\end{equation}
 and the completeness relation 
\begin{equation}
\sum_{\alpha}\tilde{\projector}_{\alpha}^{\mathcal{S}}=\idmatrix^{\mathcal{S}},\label{eq:ObservableProjectorsComplete}
\end{equation}
 where $\idmatrix^{\mathcal{S}}$ is the identity matrix for the subject
system. These eigenprojectors therefore constitute a \emph{projection-valued measure (PVM)}.
Letting $\tilde{e}_{\alpha}^{\mathcal{S}}$ be the corresponding orthonormal
basis, one has 
\begin{equation}
\tilde{e}_{\alpha}^{\mathcal{S}\adj}\tilde{e}_{\alpha^{\prime}}^{\mathcal{S}}=\delta_{\alpha\alpha^{\prime}},\qquad\tilde{e}_{\alpha}^{\mathcal{S}}\tilde{e}_{\alpha}^{\mathcal{S}\adj}=\tilde{\projector}_{\alpha}^{\mathcal{S}}.\label{eq:ObservableBasisOrthonormalComplete}
\end{equation}

If $\tilde{A}^{\mathcal{S}}$ happens to be one of the subject system's
random variables, or beables, then the eigenvalues $\tilde{a}_{\alpha}$
are its magnitudes, and the eigenprojectors $\tilde{\projector}_{\alpha}^{\mathcal{S}}$
are the system's configuration projectors. If $\tilde{A}^{\mathcal{S}}$
is instead an emergeable, then $\tilde{a}_{\alpha}$ and $\tilde{\projector}_{\alpha}^{\mathcal{S}}$
do not \emph{yet} have obvious physical meanings.

Suppose that the measuring device $\mathcal{D}$ has configurations
$d\parens{\alpha}$ that can be labeled by the same index $\alpha$
that appears in the spectral decomposition for $\tilde{A}^{\mathcal{S}}$.
Similarly, suppose that the environment $\mathcal{E}$ has configurations
$e\left(\alpha\right)$ that can also be labeled by $\alpha$.

Generalizing the unistochastic matrix \eqref{eq:CompositeSubjectEnvironmentUnistochasticMatrixFromAbsValSquareUnitary}
from the earlier analysis of the decoherence process discussed in
Subsection~\ref{subsec:Division-Events-and-the-Markov-Approximation},
suppose, moreover, that the composite system $\mathcal{S}\mathcal{D}\mathcal{E}$
evolves according to an overall unistochastic transition matrix 
\begin{equation}
\stochasticmatrix_{ide,i_{0}d_{0}e_{0}}^{\mathcal{S}\mathcal{D}\mathcal{E}}\left(t\from0\right)=\verts{\timeevop_{ide,i_{0}d_{0}e_{0}}^{\mathcal{S}\mathcal{D}\mathcal{E}}\left(t\from0\right)}^{2}.\label{eq:CompositeSubjectDeviceEnvironmentUnistochasticMatrixFromAbsValSquareUnitary}
\end{equation}
 Generalizing also the composite-system wave function \eqref{eq:CompositeSubjectEnvironmentIntermedWaveFunctionCorrelated}
from Subsection~\ref{subsec:Division-Events-and-the-Markov-Approximation},
and letting $\tilde{e}_{\alpha^{\prime},i^{\prime}}^{\mathcal{S}}$
denote the $i^{\prime}$th component of the basis vector $\tilde{e}_{\alpha^{\prime}}^{\mathcal{S}}$
with respect to the subject system's configuration basis $e_{i^{\prime}}^{\mathcal{S}}$,
suppose that the three subsystems interact up to a time $t^{\prime}>0$
in such a way that they end up with the overall wave function\footnote{It is straightforward to write down idealized examples of suitable
unitary time-evolution operators for the composite system. One choice
is $\timeevop^{\mathcal{S}\mathcal{D}\mathcal{E}}\left(t^{\prime}\right)\defeq\sum_{\alpha^{\prime}}\tilde{\projector}_{\alpha^{\prime}}^{\mathcal{S}}\tensorprod R_{d\parens{\alpha^{\prime}}}^{\mathcal{D}}\tensorprod R_{e\parens{\alpha^{\prime}}}^{\mathcal{E}}$,
where $\tilde{\projector}_{\alpha^{\prime}}^{\mathcal{S}}$ is the
$\alpha^{\prime}$th eigenprojector appearing in the spectral decomposition
for $\tilde{A}^{\mathcal{S}}$, and where $R_{d\parens{\alpha^{\prime}}}^{\mathcal{D}}$
and $R_{e\parens{\alpha^{\prime}}}^{\mathcal{E}}$ are unitary transformations
for the measuring device and the environment, respectively, that 
put them in the configurations $d\parens{\alpha^{\prime}}$ and $e\parens{\alpha^{\prime}}$.} 
\begin{equation}
\Psi_{i^{\prime}d^{\prime}e^{\prime}}^{\mathcal{S}\mathcal{D}\mathcal{E}}\left(t^{\prime}\right)=\sum_{\alpha^{\prime}}\tilde{\Psi}_{\alpha^{\prime}}^{\mathcal{S}}\left(t^{\prime}\right)\tilde{e}_{\alpha^{\prime},i^{\prime}}^{\mathcal{S}}\delta_{d^{\prime}d\parens{\alpha^{\prime}}}\delta_{e^{\prime}e\parens{\alpha^{\prime}}}.\label{eq:CompositeSubjectDeviceEnvironmentIntermedJointWaveFunctionCorrelated}
\end{equation}
 Mirroring the analogous formula \eqref{eq:CompositeSubjectEnvironmentTimeEvOpFactorizesAfterInteraction}
in  Subsection~\ref{subsec:Division-Events-and-the-Markov-Approximation},
the composite system's relative time-evolution operator factorizes
between the three subsystems for later times $t>t^{\prime}$: 
\begin{equation}
\timeevop^{\mathcal{S}\mathcal{D}\mathcal{E}}\left(t\from t^{\prime}\right)=\timeevop^{\mathcal{S}}\left(t\from t^{\prime}\right)\tensorprod\timeevop^{\mathcal{D}}\left(t\from t^{\prime}\right)\tensorprod\timeevop^{\mathcal{E}}\left(t\from t^{\prime}\right).\label{eq:CompositeSubjectDeviceEnvironmentTimeEvOpFactorizesAfterInteraction}
\end{equation}
 Then the composite system's wave function for times $t>t^{\prime}$
after the interaction is 

\begin{equation}
\begin{aligned}\Psi_{ide}^{\mathcal{S}\mathcal{D}\mathcal{E}}\left(t\right) & =\sum_{i^{\prime},e^{\prime},d^{\prime}}\timeevop_{ide,i^{\prime}d^{\prime}e^{\prime}}^{\mathcal{S}\mathcal{D}\mathcal{E}}\left(t\from t^{\prime}\right)\Psi_{i^{\prime}d^{\prime}e^{\prime}}^{\mathcal{S}\mathcal{D}\mathcal{E}}\left(t^{\prime}\right)\\
 & =\sum_{i^{\prime}}\sum_{\alpha^{\prime}}\timeevop_{ii^{\prime}}^{\mathcal{S}}\left(t\from t^{\prime}\right)\tilde{\Psi}_{\alpha^{\prime}}^{\mathcal{S}}\left(t^{\prime}\right)\tilde{e}_{\alpha^{\prime},i^{\prime}}^{\mathcal{S}}\timeevop_{dd\parens{\alpha^{\prime}}}^{\mathcal{D}}\left(t\from t^{\prime}\right)\timeevop_{ee\parens{\alpha^{\prime}}}^{\mathcal{E}}\left(t\from t^{\prime}\right).
\end{aligned}
\label{eq:CompositeSubjectDeviceEnvironmentFinalWaveFunctionCalculation}
\end{equation}

Invoking the Born rule \eqref{eq:ConfigurationBornRuleFromStateVector}
in  Subsection~\ref{subsec:The-Hilbert-Space-Representation}, it
follows from this explicit expression for the composite system's wave
function that the joint probabilities for $t>t^{\prime}$ are given
by 
\begin{equation}
p_{ide}^{\mathcal{S}\mathcal{D}\mathcal{E}}\left(t\right)=\verts{\Psi_{ide}^{\mathcal{S}\mathcal{D}\mathcal{E}}\left(t\right)}^{2}.\label{eq:CompositeSubjectDeviceEnvironmentFinalJointProbabilitiesFromAbsValSq}
\end{equation}
 Marginalizing over the configuration $i$ of the subject system $\mathcal{S}$
as well as the configuration $e$ of the environment $\mathcal{E}$,
and invoking the unitarity of both the subject system's relative time-evolution
operator $\timeevop^{\mathcal{S}}\left(t\from t^{\prime}\right)$
and the environment's relative time-evolution operator $\timeevop^{\mathcal{E}}\left(t\from t^{\prime}\right)$,
if follows from a short calculation that the standalone probabilities
$p_{d}^{\mathcal{D}}\left(t\right)$ for the measuring device $\mathcal{D}$
alone for $t>t^{\prime}$ are given by 
\begin{equation}
p_{d}^{\mathcal{D}}\left(t\right)=\sum_{\alpha^{\prime}}\verts{\timeevop_{dd\parens{\alpha^{\prime}}}^{\mathcal{D}}\left(t\from t^{\prime}\right)}^{2}\verts{\tilde{\Psi}_{\alpha^{\prime}}^{\mathcal{S}}\left(t^{\prime}\right)}^{2}.\label{eq:SubjectFinalStandaloneProbabilitiesAfterMeasurementCalculation}
\end{equation}

In the limit $t\to t^{\prime}$, this last result implies that 
\begin{equation}
p_{d\parens{\alpha^{\prime}}}^{\mathcal{D}}\left(t^{\prime}\right)=\verts{\tilde{\Psi}_{\alpha^{\prime}}^{\mathcal{S}}\left(t^{\prime}\right)}^{2}.\label{eq:DeviceBornRule}
\end{equation}
 Hence, the measuring device $\mathcal{D}$ has a standalone probability
$\verts{\tilde{\Psi}_{\alpha^{\prime}}^{\mathcal{S}}\left(t^{\prime}\right)}^{2}$
of ending up in its configuration $d\parens{\alpha^{\prime}}$, exactly
as predicted by the textbook version of the Born rule. One can then
naturally define an expectation value $\angs{\tilde{A}^{\mathcal{S}}\left(t^{\prime}\right)}$
for $\tilde{A}^{\mathcal{S}}$ at $t^{\prime}$ as the usual kind
of statistical average over device readings: 
\begin{equation}
\angs{\tilde{A}^{\mathcal{S}}\left(t^{\prime}\right)}\defeq\sum_{\alpha}\tilde{a}_{\alpha}p_{d\parens{\alpha^{\prime}}}^{\mathcal{D}}\left(t^{\prime}\right).\label{eq:ExpectationValueObservableFromMeasurementProbabilities}
\end{equation}

This analysis establishes that as long as there exists a form of unistochastic
time evolution for the composite system $\mathcal{S}\mathcal{D}\mathcal{E}$
that arrives at the appropriate wave function, the matrix $\tilde{A}^{\mathcal{S}}$
represents a genuine observable, in the sense that the time evolution
leads to the measuring device ending up in the correct outcome-configuration
with the correct Born-rule probability.

For times $t>t^{\prime}$ after the interaction, \eqref{eq:SubjectFinalStandaloneProbabilitiesAfterMeasurementCalculation}
implies that the time $t^{\prime}$ is a division event for the measuring
device, as defined in Subsection~\ref{subsec:Division-Events-and-the-Markov-Approximation}:
\begin{equation}
\stochasticmatrix^{\mathcal{D}}\left(t\right)=\stochasticmatrix^{\mathcal{D}}\left(t\from t^{\prime}\right)\stochasticmatrix^{\mathcal{D}}\left(t^{\prime}\right)\quad\textrm{for }t>t^{\prime}.\label{eq:DeviceDivisionEvent}
\end{equation}
 Here, the measuring device's dynamics for times $t>t^{\prime}$
is given by the relative unistochastic transition matrix 
\begin{equation}
\stochasticmatrix_{dd\parens{\alpha^{\prime}}}^{\mathcal{D}}\left(t\from t^{\prime}\right)\defeq\verts{\timeevop_{dd\parens{\alpha^{\prime}}}^{\mathcal{D}}\left(t\from t^{\prime}\right)}^{2}.\label{eq:DeviceRelativeUnistochasticMatrix}
\end{equation}

If the observable $\tilde{A}^{\mathcal{S}}$ is an emergeable, as
opposed to one of the subject system's (diagonal) random variables,
or beables, \eqref{eq:DefDiagonalRandomVariableMatrix}, then the
subject system $\mathcal{S}$ does not experience a division event
at $t^{\prime}$, in contrast with the measuring device $\mathcal{D}$.
Instead, the subject system remains mired in indivisible time evolution
at $t^{\prime}$, with some stochastically evolving underlying configuration.
Moreover, if indeed $\tilde{A}^{\mathcal{S}}$ is an emergeable, then
the measurement result obtained by the measuring device is an emergent
effect of the interaction between the subject system and the measuring
device rather than transparently revealing a physical aspect of
the configuration of the subject system alone. 

Despite $t^{\prime}$ not necessarily being a division event for the
subject system $\mathcal{S}$, one can nevertheless compute the standalone
probability $p_{i}^{\mathcal{S}}\left(t\right)$ for the subject system
to be in its $i$th configuration for times $t>t^{\prime}$ by marginalizing
over the measuring device $\mathcal{D}$ and the environment $\mathcal{E}$.
By another straightforward calculation, one finds 

\begin{equation}
p_{i}^{\mathcal{S}}\left(t\right)=\sum_{\alpha^{\prime}}\left[\sum_{i_{1}^{\prime},i_{2}^{\prime}}\overconj{\timeevop_{ii_{1}^{\prime}}^{\mathcal{S}}\left(t\from t^{\prime}\right)}\timeevop_{ii_{2}^{\prime}}^{\mathcal{S}}\left(t\from t^{\prime}\right)\tilde{e}_{\alpha^{\prime},i_{2}^{\prime}}^{\mathcal{S}}\overconj{\tilde{e}_{\alpha^{\prime},i_{1}^{\prime}}^{\mathcal{S}}}\right]\verts{\tilde{\Psi}_{\alpha^{\prime}}^{\mathcal{S}}\left(t^{\prime}\right)}^{2}.\label{eq:SubjectFinalStandaloneProbabilitiesFromHybridIntermedDeviceCalculation}
\end{equation}

Recognizing $\verts{\tilde{\Psi}_{\alpha^{\prime}}^{\mathcal{S}}\left(t^{\prime}\right)}^{2}$
from \eqref{eq:DeviceBornRule} as the standalone probability $p_{d\parens{\alpha^{\prime}}}^{\mathcal{D}}\left(t^{\prime}\right)$
for the measuring device $\mathcal{D}$ to end up in its configuration
$d\parens{\alpha^{\prime}}$ at the time $t^{\prime}$, and recalling
both the configuration projectors $\projector_{i}^{\mathcal{S}}$
and  the eigenprojectors $\tilde{\projector}_{\alpha}^{\mathcal{S}}$
appearing in the spectral decomposition \eqref{eq:ObservableSpectralDecomposition}
for $\tilde{A}^{\mathcal{S}}$, one can write $p_{i}^{\mathcal{S}}\left(t\right)$
more succinctly as 
\begin{equation}
p_{i}^{\mathcal{S}}\left(t\right)=\tr\parens{\projector_{i}^{\mathcal{S}}\densitymatrix^{\mathcal{S}}\left(t\right)}.\label{eq:SubjectFinalStandaloneProbabilitiesFromTraceAfterMeasurement}
\end{equation}
 Here, the subject system's density matrix $\densitymatrix^{\mathcal{S}}\left(t\right)$
for $t>t^{\prime}$ is given by 
\begin{equation}
\densitymatrix^{\mathcal{S}}\left(t\right)\defeq\timeevop^{\mathcal{S}}\left(t\from t^{\prime}\right)\left[\sum_{\alpha^{\prime}}p_{d\parens{\alpha^{\prime}}}^{\mathcal{D}}\left(t^{\prime}\right)\tilde{\projector}_{\alpha^{\prime}}^{\mathcal{S}}\right]\timeevop^{\mathcal{S}\adj}\left(t\from t^{\prime}\right).\label{eq:SubjectFinalDensityMatrixAfterMeasurement}
\end{equation}
 One can therefore recast the expectation value \eqref{eq:ExpectationValueObservableFromMeasurementProbabilities}
for $\tilde{A}^{\mathcal{S}}$ as 
\begin{equation}
\angs{\tilde{A}^{\mathcal{S}}\left(t^{\prime}\right)}=\tr\parens{\tilde{A}^{\mathcal{S}}\densitymatrix^{\mathcal{S}}\left(t^{\prime}\right)},\label{eq:ExpectationValueObservableFromTrace}
\end{equation}
 which precisely mirrors the formula \eqref{eq:ExpectationValueRandomVariableFromTrace}
for the expectation value of a (diagonal) random variable from Subsection~\ref{subsec:The-Hilbert-Space-Representation}.

Furthermore, the formula \eqref{eq:SubjectFinalStandaloneProbabilitiesFromHybridIntermedDeviceCalculation}
for $p_{i}^{\mathcal{S}}\left(t\right)$ above yields a \emph{linear}
relationship between the standalone probabilities $p_{d\parens{\alpha^{\prime}}}^{\mathcal{D}}\left(t^{\prime}\right)$
for the measuring device $\mathcal{D}$ at $t^{\prime}$ and the standalone
probabilities $p_{i}^{\mathcal{S}}\left(t\right)$ for the subject
system $\mathcal{S}$ at $t>t^{\prime}$: 
\begin{equation}
p_{i}^{\mathcal{S}}\left(t\right)=\sum_{\alpha^{\prime}}\stochasticmatrix_{i,d\parens{\alpha^{\prime}}}^{\mathcal{S}\mathcal{D}}\left(t\from t^{\prime}\right)p_{d\parens{\alpha^{\prime}}}^{\mathcal{D}}\left(t^{\prime}\right).\label{eq:SubjectFinalStandaloneProbabilitiesFromHybridIntermedDevice}
\end{equation}
 The entries $\stochasticmatrix_{i,d\parens{\alpha^{\prime}}}^{\mathcal{S}\mathcal{D}}\left(t\from t^{\prime}\right)$
of the \emph{hybrid} relative transition matrix appearing here are
given explicitly by 
\begin{equation}
\stochasticmatrix_{i,d\parens{\alpha^{\prime}}}^{\mathcal{S}\mathcal{D}}\left(t\from t^{\prime}\right)\defeq\sum_{i_{1}^{\prime},i_{2}^{\prime}}\overconj{\timeevop_{ii_{1}^{\prime}}^{\mathcal{S}}\left(t\from t^{\prime}\right)}\timeevop_{ii_{2}^{\prime}}^{\mathcal{S}}\left(t\from t^{\prime}\right)\tilde{e}_{\alpha^{\prime},i_{2}^{\prime}}^{\mathcal{S}}\overconj{\tilde{e}_{\alpha^{\prime},i_{1}^{\prime}}^{\mathcal{S}}}.\label{eq:SubjectDeviceHybridStochasticMatrix}
\end{equation}
 Because these matrix entries do not depend on the measuring device's
standalone probabilities $p_{d\parens{\alpha^{\prime}}}^{\mathcal{D}}\left(t^{\prime}\right)$,
they naturally serve as \emph{conditional probabilities} for the
subject system $\mathcal{S}$ to be in its $i$th configuration at
the time $t>t^{\prime}$, given that the measuring device $\mathcal{D}$
is in its configuration $d\parens{\alpha^{\prime}}$ at $t^{\prime}$:
\begin{equation}
p^{\mathcal{S}\mathcal{D}}\left(i,t\given d\parens{\alpha^{\prime}},t^{\prime}\right)\defeq\stochasticmatrix_{i,d\parens{\alpha^{\prime}}}^{\mathcal{S}\mathcal{D}}\left(t\from t^{\prime}\right).\label{eq:SubjectDeviceHybridConditionals}
\end{equation}

\subsection{Wave-function collapse\label{subsec:Wave-Function-Collapse}}

Importantly, notice that one can write the hybrid transition matrix
\eqref{eq:SubjectDeviceHybridStochasticMatrix} from the previous
subsection in a form that resembles the dictionary \eqref{eq:DefDictionary}:
\begin{equation}
\stochasticmatrix_{i,d\parens{\alpha^{\prime}}}^{\mathcal{S}\mathcal{D}}\left(t\from t^{\prime}\right)=\tr\parens{\timeevop^{\mathcal{S}\adj}\left(t\from t^{\prime}\right)\projector_{i}^{\mathcal{S}}\timeevop^{\mathcal{S}}\left(t\from t^{\prime}\right)\tilde{\projector}_{\alpha}^{\mathcal{S}}}.\label{eq:SubjectDeviceHybridStochasticMatrixFromDictionary}
\end{equation}
 Rearranging the right-hand side gives the equation 
\begin{equation}
\stochasticmatrix_{i,d\parens{\alpha^{\prime}}}^{\mathcal{S}\mathcal{D}}\left(t\from t^{\prime}\right)=\tr\parens{\projector_{i}^{\mathcal{S}}\densitymatrix^{\mathcal{S}\given\alpha^{\prime},t^{\prime}}\left(t\right)},\label{eq:SubjectDeviceHybridStochasticMatrixFromTraceConditionalDensityMatrix}
\end{equation}
 with a \emph{conditional} density matrix $\densitymatrix^{\mathcal{S}\given\alpha^{\prime},t^{\prime}}\left(t\right)$
for the subject system $\mathcal{S}$ at the time $t>t^{\prime}$
naturally defined by time-evolving the eigenprojector $\tilde{\projector}_{\alpha^{\prime}}^{\mathcal{S}}$
from $t^{\prime}$ to $t$: 
\begin{equation}
\densitymatrix^{\mathcal{S}\given\alpha^{\prime},t^{\prime}}\left(t\right)\defeq\timeevop^{\mathcal{S}}\left(t\from t^{\prime}\right)\tilde{\projector}_{\alpha^{\prime}}^{\mathcal{S}}\timeevop^{\mathcal{S}\adj}\left(t\from t^{\prime}\right).\label{eq:DefSubjectConditionalDensityMatrix}
\end{equation}

Thus, the calculation \eqref{eq:SubjectFinalStandaloneProbabilitiesFromHybridIntermedDeviceCalculation}
of the standalone probabilities $p_{i}^{\mathcal{S}}\left(t\right)$
for the subject system at $t>t^{\prime}$ in the previous subsection
reduces to the statement that they are given by 
\begin{equation}
p_{i}^{\mathcal{S}}\left(t\right)=\tr\parens{\projector_{i}^{\mathcal{S}}\densitymatrix^{\mathcal{S}}\left(t\right)},\label{eq:SubjectFinalStandaloneProbabilitiesFromTraceAfterMeasurementForCollapse}
\end{equation}
 where the subject system's density matrix $\densitymatrix^{\mathcal{S}}\left(t\right)$,
which was originally defined in \eqref{eq:SubjectFinalDensityMatrixAfterMeasurement}
in the previous subsection, can equivalently be expressed as a probabilistic
mixture of the conditional density matrices $\densitymatrix^{\mathcal{S}\given\alpha^{\prime},t^{\prime}}\left(t\right)$
defined in \eqref{eq:DefSubjectConditionalDensityMatrix}, statistically
weighted by the measurement probabilities $p_{d\parens{\alpha^{\prime}}}^{\mathcal{D}}\left(t^{\prime}\right)$:
\begin{equation}
\densitymatrix^{\mathcal{S}}\left(t\right)\defeq\sum_{\alpha^{\prime}}\densitymatrix^{\mathcal{S}\given\alpha^{\prime},t^{\prime}}\left(t\right)p_{d\parens{\alpha^{\prime}}}^{\mathcal{D}}\left(t^{\prime}\right).\label{eq:SubjectFindDensityMatrixFromIntermedAfterMeasurement}
\end{equation}

Taking stock of these results, one sees that to make future predictions
for $t>t^{\prime}$ about the subject system $\mathcal{S}$, conditioned
on the measuring device's result $d\parens{\alpha^{\prime}}$ at $t^{\prime}$,
one uses the conditional probabilities $\stochasticmatrix_{i,d\parens{\alpha^{\prime}}}^{\mathcal{S}\mathcal{D}}\left(t\from t^{\prime}\right)=\tr\parens{\projector_{i}^{\mathcal{S}}\densitymatrix^{\mathcal{S}\given\alpha^{\prime},t^{\prime}}\left(t\right)}$
from \eqref{eq:SubjectDeviceHybridStochasticMatrixFromTraceConditionalDensityMatrix},
in which the subject system's density matrix has effectively been
replaced by the conditional density matrix $\densitymatrix^{\mathcal{S}\given\alpha^{\prime},t^{\prime}}\left(t\right)$.
This conditional density matrix corresponds to a \emph{collapsed}
state vector or wave function defined as 
\begin{equation}
\Psi^{\mathcal{S}\given\alpha^{\prime},t^{\prime}}\left(t\right)\defeq\timeevop\left(t\from t^{\prime}\right)\tilde{e}_{\alpha}^{\mathcal{S}}.\label{eq:SubjectWaveFunctionCollapse}
\end{equation}
 The phenomenon of \emph{wave-function collapse} therefore reduces
to a prosaic example of conditioning.

By contrast, for an observer who does not know the specific measurement
result $d\parens{\alpha^{\prime}}$, the correct density matrix $\densitymatrix^{\mathcal{S}}\left(t\right)$
to use is the one defined in \eqref{eq:SubjectFindDensityMatrixFromIntermedAfterMeasurement}
in the previous subsection. Again, this density matrix consists of
an appropriate probabilistic mixture of conditional or collapsed density
matrices that are statistically weighted over the measurement results.

\subsection{The measurement problem\label{subsec:The-Measurement-Problem}}

According to the foregoing treatment of the measurement process, a
measuring device is an ordinary physical system that can carry out
a measurement of an observable and then ends up in a final configuration
that reflects a definite measurement outcome. The probabilities for
a measuring device's various possible measurement outcomes are given
by the textbook Born rule \eqref{eq:DeviceBornRule}, and conditioning
on the specific measurement outcome leads to the textbook formula
\eqref{eq:SubjectWaveFunctionCollapse} for wave-function collapse.
Hence, this picture arguably has the resources to solve the measurement
problem (Myrvold, 2022)\nocite{Myrvold:2022piiqt}. 

The stochastic-quantum correspondence is also helpful for understanding
the measurement process in another important way. Textbook treatments
typically regard measuring devices as axiomatic  primitives, without
providing clear principles for deciding which kinds of systems merit
being called measuring devices. The approach taken toward the measurement
process in this paper not only gives a candidate resolution of the
measurement problem but also yields a natural set of criteria for
defining what counts as a good measuring device in the first place,
without the need to regard measuring devices as special among all
other systems in any truly fundamental way. Based on this approach,
one sees that a good measuring device should be a physical system
with at least as many configurations as possible outcomes for the
observable to be measured (at least up to the desired level of experimental
resolution), it should admit an overall form of dynamics that results
in the correct final correlations, and it should be in sufficiently
strong contact with a noisy environment to generate a robust division
event at the conclusion of the measurement interaction.\footnote{The first two of these three criteria would be standard requirements
for a measuring device even without worrying about indivisible stochastic
dynamics or quantum theory. Note that without the third criterion\textemdash strong
contact with an environment\textemdash one obtains a ``latent measurement''
(Dicke 1989; Glick, Adami 2020)\nocite{Dicke:1989qmsal,GlickAdami:2020manmqm}.}

\subsection{The uncertainty principle\label{subsec:The-Uncertainty-Principle}}

Again, the preceding treatment of the measurement process leads to
the textbook Born rule \eqref{eq:DeviceBornRule} and the textbook
formula \eqref{eq:SubjectWaveFunctionCollapse} for wave-function
collapse. As a consequence, any pair of observables $\tilde{A},\tilde{B}$
and their respective standard deviations $\Delta\tilde{A},\Delta\tilde{B}$
will satisfy the \emph{Heisenberg-Robertson uncertainty principle}
(Heisenberg 1927, Robertson 1929)\nocite{Heisenberg:1927udaidqkum,Robertson:1929tup},
\begin{equation}
\Delta\tilde{A}\,\Delta\tilde{B}\geq\frac{1}{2}\verts{\tr\parens{i\bracks{\tilde{A}\tilde{B}-\tilde{B}\tilde{A}}\densitymatrix}},\label{eq:HeisenbergRobertsonUncertaintyPrinciple}
\end{equation}
 as follows from any of the standard proofs.

The stochastic-quantum correspondence goes beyond replicating the
uncertainty principle by painting a clearer picture of what the uncertainty
principle physically means. Consider for simplicity the case in which
$\tilde{A}=A$ is a random variable, or beable, and $\tilde{B}$ is
an emergeable, in the language of Subsection~\ref{subsec:Emergeables}.
Then $A$ has a direct interpretation solely in terms of the subject
system's configuration space, whereas $\tilde{B}$ encodes an emergent
pattern in the subject system's dynamics that can nonetheless show
up in the measurement outcomes of a measuring device.

Suppose that $A$ has a definite value or magnitude at some initial
time $0$. Then, assuming that $A$ has no degeneracies in its spectrum,
the subject system must be in a specific configuration with probability
$1$ at the initial time $0$. The overall stochastic dynamics will
then lead to uncertainty in the outcome of any measurement of $\tilde{B}$.

Suppose that one goes ahead and measures $\tilde{B}$, so that a definite
measurement outcome emergently shows up in the configuration of a
measuring device at some time $t^{\prime}>0$. The analysis in Subsection~\ref{subsec:Wave-Function-Collapse}
then implies that there is an inevitable disturbance in the subject
system that leads its density matrix to end up effectively as a non-diagonal
matrix equal to an eigenprojector of $\tilde{B}$. A non-diagonal
density matrix signifies  that the system is in the midst of an indivisible
stochastic process, as explained in Subsection~\ref{subsec:Decoherence}.
In the present circumstances, that indivisible stochastic process
is precisely one that would ensure that if $\tilde{B}$ were measured
again shortly after $t^{\prime}$, then the measuring device would
obtain the same outcome for $\tilde{B}$ as before. However, being
in the midst of an indivisible stochastic process also implies uncertainty
in the subject system's underlying configuration, thereby rendering
the value of $A$ uncertain.

\section{Discussion and Future Work\label{sec:Discussion-and-Future-Work}}

\subsection{Indivisible quantum theory\label{subsec:Indivisible-Quantum-Theory}}

This paper has shown that one can reconstruct the mathematical formalism
and all the empirical predictions of quantum theory using simpler,
more physically transparent axioms than the standard Dirac-von Neumann
axioms. Rather than postulating Hilbert spaces and their ingredients
from the beginning, one instead posits a physical model, called an
\emph{indivisible stochastic process}, based on trajectories in configuration
spaces following generically indivisible stochastic dynamics. The
\emph{stochastic-quantum correspondence} then shows that every quantum
system can be viewed as the Hilbert-space representation of an underlying
indivisible stochastic processes.

This new axiomatic approach naturally suggests a new interpretation
of quantum theory grounded in the theory of stochastic processes.
According to this highly adaptable interpretation, which one could
naturally call the \emph{indivisible interpretation} of quantum theory,
or just \emph{indivisible quantum theory}, systems have underlying
physical configurations in configuration spaces at all times.

This perspective deflates some of the most mysterious features of
quantum theory. In particular, one sees that density matrices, wave
functions, and all the other ingredients of Hilbert spaces, while
highly useful, are merely mathematical appurtenances. These appurtenances
should therefore not be assigned direct physical meanings or treated
as though they directly represent physical objects, any more than
Lagrangians or Hamilton's principal functions directly represent physical
objects. Superposition is then not a literal smearing of physical
objects but is merely a mathematical artifact of catching a system
in the middle of an indivisible stochastic process, as represented
using a Hilbert-space formulation and wave functions.

Moreover, from this standpoint, \emph{canonical quantization} need
not be regarded as the promotion of classical observables to noncommutative
operators by fiat but can be implemented (when mathematically feasible)
simply by generalizing a classical system's dynamics from being deterministic
to being stochastic, with all the exotic features of quantum theory
then emerging automatically. As a consequence, this formulation of
canonical quantization potentially offers more straightforward techniques
for coupling classical systems to quantum systems in real-world applications.

\subsection{The category problem\label{subsec:The-Category-Problem}}

In an important sense, the stochastic-quantum correspondence and the
indivisible interpretation also legitimize many standard practices
followed in physics and in other scientific areas like astronomy,
chemistry, biology, and paleontology. To see why, notice that according
to the thoroughly instrumentalist and operationalist Dirac-von Neumann
axioms, the \emph{only} predictions provided by textbook quantum theory
are predictions about a rather narrow category of things: measurement
outcomes, probabilities of measurement outcomes, and expectation values
that are averages of measurement outcomes statistically weighted by
measurement-outcome probabilities (Griffiths 2018; Townsend 2012;
Shankar 1994; Sakurai, Napolitano 2010; Schumacher, Westmoreland 2010)\nocite{GriffithsSchroeter:2018iqm,Townsend:2012maqm,Shankar:1994pqm,SakuraiNapolitano:2010mqm,SchumacherWestmoreland:2010qpsi}.
Meanwhile, scientists in all areas of research talk about a much broader
category of phenomena\textemdash from the mixing of gases in the primordial
universe to the spontaneous appearance of genetic mutations\textemdash that
presumably just \emph{happen} in some way, according to \emph{happening}
probabilities, in the past, present, or future. Strictly speaking,
however, the \emph{happening} of phenomena, as a category, lies outside
the axiomatic ambit of textbook quantum theory, which refers only
to connecting the measurement settings of chemical detectors and telescopes
to the probabilities of their measurement outcomes. The inability
of textbook quantum theory to account for the happening of phenomena
represents what one might call the \emph{category problem}. The category
problem either means that scientists are not speaking honestly or
coherently about their research, or that textbook quantum theory is
inadequate as a physical theory.

Decoherence alone cannot bridge the categorical gap between measurement-outcome
probabilities and happening probabilities, because decoherence can
only temporarily change whatever orthonormal basis momentarily diagonalizes
a system's density matrix (and, after all, every density matrix is
always diagonal in \emph{some} orthonormal basis). After a system
undergoes decoherence, textbook quantum theory then still requires
one to make a direct appeal to the measurement axioms to translate
the final density matrix into a statement about probabilities, which
will then axiomatically end up being measurement-outcome probabilities
rather than happening probabilities.

Nor can appealing to some sort of \emph{thermodynamic limit} resolve
the discrepancy either. In order for a limit in a physical context
to make sense, there should be clearly physical ingredients or constituents
involved. Furthermore, the end result of the limit should gradually
emerge as a better and better physical approximation at \emph{finite}
stages of the limiting process, simply because a rigorous limit consists
of inequalities between finite (if arbitrarily large or small) parameters.
For example, in the \emph{hydrodynamic limit} of a system of classical
interacting particles, the particles are the physical ingredients,
and one sees fluid-like behavior gradually emerge as a better and
better physical approximation as the number of particles progressively
increases. In the case of textbook quantum theory, by contrast, every
finite stage of any purported thermodynamic limit features only measurement
outcomes and measurement-outcome probabilities, so there are no clearly
physical ingredients or constituents, and the categorical gap between
measurement outcomes and the happening of phenomena never closes.

The stochastic-quantum correspondence and the indivisible interpretation
yield a much richer version of quantum theory in which physical phenomena
really happen, with probabilities that are really happening probabilities,
and therefore not only resolves the category problem but also vindicates
the ways that scientists talk about the world. Measurement-outcome
probabilities are then merely a special case, arising when what is
actually happening is a change to the configuration of a measuring
device.

\subsection{Interpretational issues\label{subsec:Interpretational-issues}}

A formulation or interpretation of quantum theory that posits physical
configurations separate from\textemdash or, in the present case, instead
of\textemdash the standard ingredients of Hilbert spaces can be thought
of as a kind of \emph{hidden-variables theory}\index{hidden-variables theory}.
In keeping with the Bell-Kochen-Specker theorem (Bell 1966; Kochen,
Specker 1967)\nocite{Bell:1966otpohviqm,KochenSpecker:1967phvqm},
indivisible quantum theory is a manifestly contextual theory, with
a given quantum system's beables belonging to a specific measurement
context, and various classes of emergeables belonging to other measurement
contexts, as detailed in Subsection~\ref{subsec:The-Measurement-Process}.

Indivisible quantum theory is based on non-Markovian stochastic dynamics,
so it lies outside the \emph{ontological models} framework of Harrigan
and Spekkens (2010)\nocite{HarriganSpekkens:2010eievqs}. In particular,
the wave function is neither ontic nor entirely epistemic but has
a law-like or nomic character, as is clear from its definition \eqref{eq:DefStateVector}
as part of the time-evolution operator.\footnote{Indeed, in Section 6 of their 2010 paper, Harrigan and Spekkens specifically
note that models based on ``nomic'' wave functions and stochastic
dynamical laws lie outside their framework.} As such, the theorem of Pusey, Barrett, and Rudolph (2012) \nocite{PuseyBarrettRudolph:2012rqs}
does not apply.

Because indivisible quantum theory  invokes hidden variables in the
form of underlying physical configurations, this framework for quantum
theory shares some aspects with the \emph{de Broglie-Bohm formulation},
or \emph{Bohmian mechanics}~(de Broglie 1930; Bohm 1952a, 1952b)\nocite{deBroglie:1930iswm,Bohm:1952siqtthvi,Bohm:1952siqtthvii}.
However, in contrast to indivisible quantum theory, Bohmian mechanics
employs deterministic dynamics and features a fundamental guiding
equation that explicitly breaks Lorentz invariance by singling out
a preferred foliation of spacetime into spacelike hypersurfaces. The
indivisible interpretation instead takes seriously what experiments
strongly suggest\textemdash that the dynamics of quantum theory is
indeterministic, that there is no fundamentally preferred foliation
of spacetime, and that quantum systems can exhibit genuinely non-Markovian
behavior (Glick, Adami 2020)\nocite{GlickAdami:2020manmqm}. Indivisible
quantum theory is also more flexible and model-independent than Bohmian
mechanics and works for all kinds of quantum systems, beyond the
case of systems of fixed numbers of finitely many non-relativistic
particles.

In contrast with the \emph{Everett interpretation} (Everett 1957,
1973; Wallace 2012)\nocite{Everett:1957rsfqm,Everett:1973tuwf,Wallace:2012temqtattei},
also known as the \emph{\textquoteleft many worlds\textquoteright{} interpretation},
the indivisible interpretation assumes that quantum systems, like
classical systems, have definite configurations in configuration spaces
and does not attempt to derive probability from non-probabilistic
assumptions or grapple with fundamental aspects of personal identity
in a universe continuously branching into large (and somewhat undefined)
numbers of parallel worlds. Simply put, there are no fundamental wave
functions in the indivisible interpretation, meaning that there is
nothing in the ontology that branches into a multitude of worlds,
so the approach taken in this paper is more metaphysically modest
than the Everett interpretation.

Unlike \emph{stochastic-collapse theories} (Ghirardi, Remini, Weber
1986; Bassi, Ghirardi 2003)\nocite{GhirardiRiminiWeber:1986udmms,BassiGhirardi:2003drm},
indivisible quantum theory does not invoke any fundamental violations
of unitarity and does not require introducing any new constants of
nature to specify dynamical-collapse rates. That said, there are common
threads between indivisible quantum theory and some approaches to
stochastic-collapse theories that demote the wave function from having
a physical status (Bedingham 2018)\nocite{Bedingham:2018cmasts}.\footnote{The author would like to thank an anonymous reviewer for suggesting
this point.}

The indivisible interpretation shares some features with the \emph{modal interpretations}
(Krips 1969; Van Fraassen 1972; Vermaas, Dieks 1995; Bacciagaluppi,
Hemmo 1996; Lombardi, Dieks 2021)\nocite{Krips:1969tpqm,vanFraassen:1972afattpos,VermaasDieks:1995miqmgdo,BacciagaluppiHemmo:1996midam,LombardiDieks:2021mioqm},
including an insistence that systems always have definite configurations
of some kind at every moment in time, while assigning a law-like,
objective role to at least some forms of probability.  One difference
between the indivisible interpretation and most of the modal interpretations,
however, is the indivisible interpretation's insistence that the definite
configuration of a given system is an element of a classical-looking
configuration space rather than corresponding more abstractly to
features of a Hilbert space. The indivisible interpretation also avoids
some of the ontological instabilities that are a serious challenge
for most of the modal interpretations (Vermaas 1999)\nocite{Vermaas:1999apuoqmpaioami}.

\subsection{Future directions\label{subsec:Future-Directions}}

Future work will address questions of locality and causation. Looking
forward, it would also be interesting to see what implications the
stochastic-quantum correspondence could have for both phenomenological
stochastic processes, like those in biology or finance, as well as
for future work in fundamental physics, like quantum gravity.

More broadly, by recasting the Hilbert-space formulation of quantum
theory as merely a convenient way to represent a large class of stochastic
processes, one opens the door to searching for totally different representations
that might look nothing at all like Hilbert spaces and that could
allow for the construction of more general kinds of stochastic processes.
Perhaps one could even find a way to generalize the theory beyond
stochastic processes altogether.

\section*{Acknowledgments}

The author would especially like to acknowledge Emily Adlam, David
Albert, Howard Georgi, David Kagan, and Logan McCarty for extensive
discussions during the writing of this paper. The author would also
like to thank Scott Aaronson, Ignacio Cirac, Ned Hall, David Kaiser,
Serhii Kryhin, Barry Loewer, Alex Meehan, Xiao-Li Meng, Simon Milz,
Kavan Modi, Wayne Myrvold, Filip Niewinski, Jill North, Jamie Robins,
Noel Swanson, Vivishek Sudhir, Xi Yin, and Nicole Yunger Halpern for
helpful conversations.

\section*{Declarations}

\textbf{Funding, Conflicts of Interest, and Data Availability:} The
author has no funding sources to declare, any conflicts of interest
to report, or any data to make available.

\bibliographystyle{1_home_jacob_Documents_Work_My_Papers_2023-Stoc___ses_and_Quantum_Theory_custom-abbrvalphaurl}
\bibliography{0_home_jacob_Documents_Work_My_Papers_Bibliography_Global-Bibliography}

\end{document}